\documentclass[11pt,a4paper,oneside]{article}

\usepackage{amsmath}%
\usepackage{amsfonts}%
\usepackage{amssymb}%
\usepackage{adjustbox}%
\usepackage{graphicx}%
\usepackage{graphics,color}%
\usepackage{natbib}%
\usepackage{subfigure}

\newtheorem{theorem}{Theorem}

\begin{document}
\title{Model Prior Distribution for Variable Selection in Linear Regression Models}

\author{ {\large Cristiano Villa}\footnote{{\sc University of Kent, School of Mathematics, Statistics and Actuarial Sciences, Canterbury, UK} E-mail: cv88@kent.ac.uk (corresponding author)} \,{\large and} {\large Jeong Eun Lee}\footnote{{\sc Auckland University of Technology, School of Computing and Mathematical Sciences, Auckland, New Zealand} E-mail:  jeong.lee@aut.ac.nz} 
\\
}
\maketitle

\begin{abstract}
In this work we discuss a novel model prior probability for variable selection in linear regression. The idea is to determine the prior mass in an objective sense, by considering the \emph{worth} of each of the possible regression models, given the number of covariates under consideration. Through a simulation study, we show that the proposed prior outperforms the uniform prior and the Scott \& Berger prior in a scenario of no prior knowledge about the size of the true regression models. We illustrate the use of the prior using two well-known data sets with, respectively, 15 and 4 covariates.
\end{abstract}

\noindent \textbf{Keywords} Bayesian variable selection, linear regression, loss functions, objective priors.

%%-----------------------------------------------------------------------------------------------
%--- INTRODUCTION -------------------------------------------------------------------------------
\section{Introduction}\label{sc_intro}
In this paper, we propose a model prior probability distribution for variable selection problems in linear regression models. We focus on the general approach where, given $d$ covariates, the task is to derive a posterior probability on the space of regression models. The prior we propose is based on losses and it is compared with both the uniform prior (which assigns equal prior mass to each model) and the Scott \& Berger prior \citep{SB10} through a simulation study. We show that, in absence of any prior information about the true regression model, the model prior we propose has an overall better performance than the other two. We also compare the model prior distributions on two well-known data sets: the US crime data \citep{V78} and the Hald data \citep{Woods32}.

Variable selection problems, in the Bayesian framework, are (and should be) in line with any other inferential procedure. That is, a posterior distribution for the space of models should be obtained in order to represent the (posterior) uncertainty about the true regression model (see \cite{Gelman04}). There may be instances where the above is not appropriate, for example if there are models with a negligible posterior probability, in which case a subset of all the possible regression models can be considered.
With a prior distribution on the space of models, representing the model uncertainty related to variable selection, one way to proceed is by using Bayesian model averaging \citep{Hoet99}. When the model posterior distribution tends to be spread across many of the possible regression models, and when prediction is an important part of the statistical analysis, \cite{Rafte97} show that Bayesian model averaging performs superiorly than choosing the regression model with the highest posterior probability. Also, although one may decide to explore a different route than model averaging, \cite{BB04} show that the median probability model, under certain conditions, has a predictive power at least as good as the one of the highest probability model. The median probability model derives from an equivalent ``answer'' to the above problem obtained by estimating the marginal posterior probability that each variable has, independently from the others, of being included in the regression model.

An important component of the Bayesian variable selection approach is the definition of the prior for the regression coefficients, including the intercept, and the regression variance. In fact, we can comfortably say that the majority of literature related to variable selection is focussed on the identification of appropriate prior distributions for the model specific parameters. This is particularly the case when the approach has to be minimally informative; however, due to the large number of potential covariates examined in modern problems, a minimally informative prior is possibly the most wise (or the sole) solution as it would be not feasible to elicit priors for moderate to large models. Although the importance of parameter specific priors, in this paper we will focus on the model prior distribution only, referring to the specific literature on the subject. See, for example, \cite{Betal12} and the references therein.

The prior we propose is based on the idea that, given a variable selection problem, where all the potential covariates have been identified, every possible linear regression model has a \emph{worth} depending on the fact that that model has been chosen to be part of the problem \citep{VW15}. In other words, the choice of including a model in the model space conveys information, and we use that information to assign a prior probability to the model. It is important to highlight that the prior we propose is intended to be applied within the Bayesian framework, in the sense that the prior represents the initial uncertainty about the true model which is updated, by including the information coming from the data, to obtain a posterior distribution representing the uncertainty at the end of the process \citep{Gelman04}. The obtained prior depends on the model size, that is on the number of covariates included. As such, we will compare the proposed prior with the uniform and the Scott \& Berger priors on the basis of the frequentist performances of posterior for the model size.

The paper is organised as follows. In Section \ref{sc_notation} we define the notation used throughout the paper and formalise the problem of variable selection for linear regression models in a Bayesian framework. In Section \ref{sc_priors} we discuss the current objective model priors for variable selection (the uniform prior and the Scott \& Berger prior), and we also present the proposed prior based on losses. The results of simulation studies are provided in Section \ref{sc_simulation}. In the section we examine the performance of the considered priors on the basis of the frequentist results of the corresponding model size posterior distributions. Section \ref{sc_realdata} reports the analysis results using two real data sets widely discussed in literature. The final Section \ref{sc_conclusion} concludes and provides some discussion points on the proposed prior and its comparison with the other two priors.

\section{Notation and problem specification}\label{sc_notation}
The Bayesian set up for variable selection problems is, as follows. The task is to explain a response variable by means of a set of $d$ possible covariates, with a sample of observations of size $n$. Given the vector $\mathbf{y}$ of $n$ responses, the design matrix $\mathbf{X}$ of size $n\times d$, an intercept $\alpha$ and a vector of coefficients $\pmb\beta$ of dimension $d$, the response outcome $y_i$ is expressed as
\begin{equation}\label{eq_model}
y_i = \alpha + \sum_{j=1}^d\beta_j X_{ij} + \varepsilon_i, \qquad i=1,\ldots,n,
\end{equation}
where $\varepsilon_i\sim N(0,1/\phi)$, for $i=1,\ldots,n$, are the i.i.d. normally distributed errors with unknown variance $1/\phi$. We assume that the number of observations $n$ is larger than the number of covariates (i.e. $n>d$), and the design matrix is of full rank. The variable selection problem can be seen as identifying which of the possible $d$ covariates has impact on $\mathbf{y}$. In other words, we aim to identify which of the regression parameters $\beta_j$s are different from zero. Let us consider the binary vector $\boldsymbol{\gamma}$, where the $j$-th element is zero if $\beta_j=0$ and one if $\beta_j\neq0$. Then, the generic Bayesian regression model is indicated by
\begin{equation}\label{eq_regmodel1}
M_{\pmb\gamma} = \left\{f(\mathbf{y}|\alpha,\pmb\beta_{\pmb\gamma},\phi) ; \pi_{\pmb\gamma}(\alpha,\pmb\beta_{\pmb\gamma},\phi)\right\},
\end{equation}
where
$$f(\mathbf{y}|\alpha,\pmb\beta_{\pmb\gamma},\phi) = N(\mathbf{y}|\alpha+\mathbf{X}_{\pmb\gamma}\pmb\beta_{\pmb\gamma},\mbox{I}/\phi),$$
and $\pi_{\pmb\gamma}(\alpha,\pmb\beta_{\pmb\gamma},\phi)$ represents the prior distribution for the parameters of the model, the so-called model-specific parameter prior. Note that the cardinality of $\pmb\gamma$ gives the number of covariates included in model $M_{\boldsymbol{\gamma}}$. There are $2^d$ possible regression models, each one of them identified by $\pmb\gamma$.

In the Bayesian framework, inference about model uncertainty is based on the model posterior probability
$$p(M_{\pmb\gamma}|\mathbf{y}) \propto m(\mathbf{y}|M_{\pmb\gamma})p(M_{\pmb\gamma}),$$
where $p(M_{\pmb\gamma})$ is the prior probability for model $M_{\pmb\gamma}$ and
\begin{equation}\label{eq_marglike1}
m(\mathbf{y}|M_{\pmb\gamma}) = \int f(\mathbf{y}|\alpha,\pmb\beta_{\pmb\gamma},\phi) \pi_{\pmb\gamma}(\alpha,\pmb\beta_{\pmb\gamma},\phi)\,d\alpha\,d\pmb\beta_{\pmb\gamma}\,d\phi,
\end{equation}
is the marginal likelihood of the observations under model $M_{\pmb\gamma}$. The model posterior distribution can then be used to either estimate a specific regression model or perform model averaging.

The number of possible regression models grows exponentially with $d$. When $d$ is large the posterior probabilities in \eqref{eq_marglike1} are often small for most of models, and posterior inclusion probabilities could give a better idea of the posterior uncertainty in comparison to model posterior probabilities. The posterior inclusion probability of the $j$-th covariate is defined as
$$\omega_j = \mbox{Pr}(\gamma_j\neq0|\mathbf{y}) = \sum_{\boldsymbol{\gamma}}p(M_{\boldsymbol{\gamma}}|\mathbf{y})\cdot 1_{\gamma_j=1}.$$
Prior and posterior inclusion probabilities originate from the common idea in Bayesian variable selection to consider variable inclusions as exchangeable Bernoulli trials, with the common inclusion probability $\omega$, implying
\begin{equation}\label{eq_berntrial}
p(M_{\boldsymbol{\gamma}}|\omega) = \omega^{|\boldsymbol{\gamma}|}(1-\omega)^{d-|\boldsymbol{\gamma}|}.
\end{equation}

\paragraph{Model-specific parameter prior}
Prior choice on model-specific parameters has received much attention, and well-received priors include the Zellner--Siow prior \citep{ZS80}, the Zellner's $g$-prior \citep{Z86}, the mixtures of $g$-priors \citep{Letal08}, and the more recent \emph{robust} prior by \cite{Betal12}, among others. The robust prior has the advantage of yielding closed-form marginal likelihoods and to not suffer from the information paradox \citep{Letal08}.
The robust prior is defined as
\begin{eqnarray}\label{eq_robustprior}
\pi_{\pmb\gamma}(\alpha,\pmb\beta_{\pmb\gamma},\phi) &=& \pi(\alpha,\phi)\times\pi(\pmb\beta_{\pmb\gamma}) \nonumber \\
&=& \phi^{1/2}\int_0^\infty N_{|\pmb\gamma|}(\pmb\beta_{\pmb\gamma}|0,g\Sigma_{\pmb\gamma})\pi_{\pmb\gamma}(g)\,d g,
\end{eqnarray}
where $\Sigma_{\pmb\gamma}$ is the covariance matrix of the maximum likelihood estimator of $\pmb\beta_{\pmb\gamma}$. The distribution of $g$ is given by
$$\pi_{\pmb\gamma}(g) = a[\rho_{\pmb\gamma}(b+n)]^a(g+b)^{-(a+1)} \cdot 1_{\left\{g>\rho_{\pmb\gamma}(b+n)-b\right\}},$$
where $a,b>0$ and $\rho_{\pmb\gamma}\geq b/(b+n)$. The prior \eqref{eq_robustprior} is called \emph{robust} as its tails behave like the tails of a multivariate $t$ density, therefore less sensible to outliers.

In this paper, as the focus is solely on model priors, every analysis is performed by using the same model-specific parameter prior, the robust prior with hyperparameter values $a=1/2$, $b=1$ and $\rho_{\pmb\gamma}=1/(d+1)$, as recommended in \cite{Betal12}, so differences in the results can be ascribed to differences in the model prior.

%--- MODEL PRIORS -------------------------------------------------------------------------------
\section{Model priors in objective variable selection}\label{sc_priors}
To the best of our knowledge, the choice of model prior probabilities which convey minimal information is limited to two options: the uniform prior and the so-called Scoot \& Berger prior.

The model uniform prior is obtained by assigning equal prior mass to each regression model, that is $p(M_{\pmb\gamma})=1/2^d$ for any $\pmb\gamma$, and it yields a prior inclusion probability of $\omega=1/2$.
\cite{SB10} discuss the following model prior for variable selection
\begin{eqnarray}\label{eq_sbprior}
p(M_{\pmb\gamma}) &=& \int_0^1 p(M_{\pmb\gamma}|\omega)\pi_\omega(\omega)\,d\omega \nonumber \\
&=& \frac{1}{d+1}\dbinom{d}{|\gamma|}^{-1}.
\end{eqnarray}
Their model prior is obtained by assigning a beta prior to $\omega$, with both the hyper-parameters equal to one, and then marginalising over $\omega$. This prior was previously discussed in the literature, see for example \cite{LS09}, with the aim of representing prior minimal information. The motivation behind the choice of the model prior by \cite{SB10} lies in its property to correct for multiplicity, which can be seen as an issue when model choice is performed by multiple statistical testing with respect to a reference model (typically the null model or the full model).
The Scott \& Berger prior induces a marginal prior inclusion probability of $\omega=1/2$ for each covariate, same as the uniform model prior. However, as thoroughly discussed in \cite{SB10}, given that their prior is function of $d$, it allows the multiplicity correction. It may also be worthwhile to mention that the choice of a uniform prior for $\omega$, which is a probability of success of Bernoulli trials, somehow contradicts the usual choice of a $\mbox{Beta}(1/2,1/2)$ as default prior in parameter inference \citep{Jef61,BS94}.

\subsection{Model prior based on losses}
The model prior based on losses has been introduced by \cite{VW15}. The basic idea is that we can assign a \emph{worth} to each model by objectively measuring what is lost if the model is removed from the space of models, and it is the true one.

To derive the prior for $M_{\pmb\gamma}$ we then follow the framework detailed in \cite{VW15}. First, we derive the loss in information when the true model is removed, which we indicate by $\mbox{L}_I(M_{\pmb\gamma})$. We apply the well-known Bayesian result in \cite{Berk66}, which states that if a model is misspecified the posterior asymptotically accumulates on the nearest model, where the nearest model is the one which minimises the Kullback--Leibler divergence \citep{KL51} from the true model. To illustrate, let us assume that we want to assess the \emph{worth} of the regression model $M_{\pmb\gamma}$, with the simplified notation $\pmb\theta_{\pmb\gamma}=(\alpha,\pmb\beta_{\pmb\gamma},\phi)$. The loss in information by removing model $M_{\boldsymbol{\gamma}}$, when it is the true model, is given by
$$L_I(M_{\pmb\gamma}) = - \int_{\pmb\theta_{\pmb\gamma}}\inf_{\pmb\theta_{\pmb\gamma^\prime}\neq \pmb\theta_{\pmb\gamma}}D_{KL}\Big(f(\mathbf{y}|\pmb\theta_{\pmb\gamma})\|f(\mathbf{y}|\pmb\theta_{\pmb\gamma^\prime})\Big)\pi_{\pmb\gamma}(\pmb\theta_{\pmb\gamma})\,d\pmb\theta_{\pmb\gamma},$$
where $f(\mathbf{y}|\pmb\theta_{\pmb\gamma^\prime})$ represents the regression distribution of $M_{\pmb\gamma^\prime}$, that is, the regression model which is the most similar to $f(\mathbf{y}|\pmb\theta_{\pmb\gamma})$. We then see that the loss in information associated to model $M_{\pmb\gamma}$ is the expected minimum Kullback--Leibler divergence between $M_{\pmb\gamma}$ and the nearest one, where the expectation is taken with respect to the prior $\pi_{\pmb\gamma}(\pmb\theta_{\pmb\gamma})$, representing the prior uncertainty about the true values of $\alpha$, $\pmb\beta_{\pmb\gamma}$ and $\phi$.

To link the \emph{worth} of a model to its prior probability, we use the \emph{self-information} loss function, as discussed in \cite{VW15}. Briefly, this particular type of loss function (also known as the \emph{log-loss} function) measures the performance of a probability statement with respect to an outcome. Thus, for every probability assignment $P=\{\mbox{Pr}(A),A\in\Omega\}$, the self-information loss function is defined as
$$\mbox{L}(P,A) = -\log \mbox{Pr}(A).$$
More details and properties of this particular loss function can be found, for example, in \cite{MF98}. As both the expected minimum Kullback--Leibler divergence between model $M_{\pmb\gamma}$ and the nearest model, and the self-information loss function measure the same quantity, i.e. the loss in information, we can equate the two measurements yielding
\begin{equation}\label{eq_priorinfo}
p(M_{\pmb\gamma}) \propto \exp\left\{\min_{\pmb\gamma^\prime\neq\pmb\gamma}D_{KL}\Big(f(\mathbf{y}|\pmb\theta_{\pmb\gamma})\|f(\mathbf{y}|\pmb\theta_{\pmb\gamma^\prime}\Big)\right\}.
\end{equation}
However, as the number of covariates may easily be moderate to large, it is also necessary to take into considerations the complexity of the model. For the regression model $M_{\pmb\gamma}$, we denote the loss due to complexity by $\mbox{L}_C(M_{\pmb\gamma})$, and it is determined as follows. If we keep model $M_{\pmb\gamma}$ in the space of models, the loss would be proportional to the number of covariates that have to be considered and measured. Therefore, the loss of keeping a linear regression model increases with the number of covariates it contains, and we have
$$\mbox{L}(\mbox{remove } M_{\pmb\gamma}) = \mbox{U}(\mbox{keep } M_{\pmb\gamma}) = -c\cdot |\pmb\gamma|,$$
and
$$\mbox{L}(\mbox{keep }M_{\pmb\gamma}) = c\cdot |\pmb\gamma|, \qquad c>0,$$
where $\mbox{L}(\cdot)$ represents a loss and $\mbox{U}(\cdot)$ a utility.

The loss component due to complexity is easily fit in our framework and the model prior for $M_{\pmb\gamma}$ is
\begin{equation}\label{eq_totalprior}
p(M_{\pmb\gamma}) \propto \exp\left\{ \int_{\pmb\theta_{\pmb\gamma}} \left[\inf_{\pmb\gamma^\prime\neq\pmb\gamma}D_{KL}\Big(f(\mathbf{y}|\pmb\theta_{\pmb\gamma})\|f(\mathbf{y}|\pmb\theta_{\pmb\gamma^\prime})\Big)\pi_{\pmb\gamma}(\pmb\theta_{\pmb\gamma})\,d\pmb\theta_{\pmb\gamma}\right] - c\cdot|\pmb\gamma| \right\}.
\end{equation}
In other words, the prior is constructed based upon a cumulative loss with a component representing the loss in information and a component representing the loss due to complexity.

The following Theorem \ref{teo_KLreg1} (which proof is in the Appendix) shows the expression of the minimum Kullback--Leibler divergence between regression models.
\begin{theorem}\label{teo_KLreg1}
Let $M_{\pmb\gamma}=\{f(\mathbf{y}|\pmb\theta_{\pmb\gamma});\pi_{\pmb\gamma}(\pmb\theta_{\pmb\gamma})\}$ and $M_{\pmb\gamma^\prime}=\{f(\mathbf{y}|\pmb\theta_{\pmb\gamma^\prime});\pi_{\pmb\gamma^\prime}(\pmb\theta_{\pmb\gamma^\prime})\}$ be linear normal regression models as in \eqref{eq_regmodel1}, with design matrices, respectively, $\mathbf{X}_{\pmb\gamma}$ and $\mathbf{X}_{\pmb\gamma^\prime}$. If $\mathbf{X}^T_{\pmb\gamma}\mathbf{X}_{\pmb\gamma^\prime}$ is invertible, the minimum Kullback--Leibler divergence between $f(\mathbf{y}|\pmb\theta_{\pmb\gamma})$ and $f(\mathbf{y}|\pmb\theta_{\pmb\gamma^\prime})$ is  
\begin{equation}\label{eq_minkltheo1}
\min_{\pmb\theta_{\pmb\gamma^\prime}} D_{KL}\Big(f(\mathbf{y}|\pmb\theta_{\pmb\gamma})\|f(\mathbf{y}|\pmb\theta_{\pmb\gamma^\prime})\Big) = 0 \hspace{1cm} \forall\, \pmb\gamma\neq\pmb\gamma^\prime.
\end{equation}
\end{theorem}
Theorem \ref{teo_KLreg1} shows that the minimum Kullback--Leibler divergence between any two linear regression models is zero, regardless to the number of covariates in the models. This means that, in variable selection problems for linear regression models, there is no loss in information in selecting the ``wrong'' model, as such the model prior in \eqref{eq_totalprior} becomes
\begin{equation}\label{eq_ourprior}
p(M_{\pmb\gamma}) \propto \exp\left\{- c\cdot|\pmb\gamma|\right\}.
\end{equation}

\begin{figure}[h]
\centering
\includegraphics[scale=0.7]{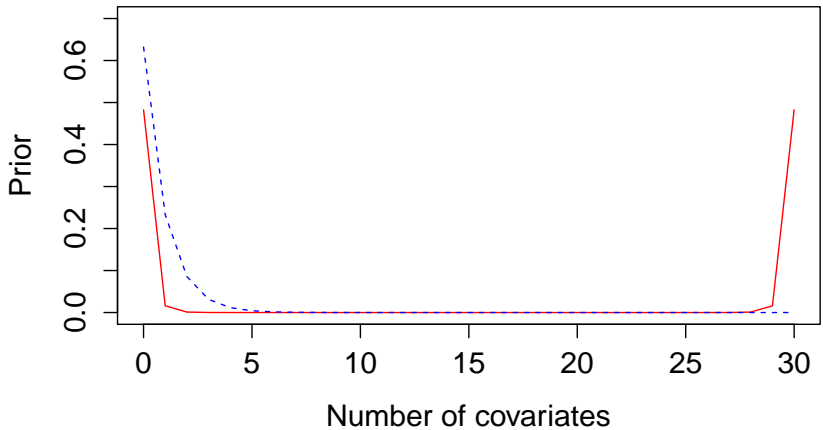}
\caption{Prior model comparison when the full model contains 30 covariates: Scott \& Berger prior (continuous red line) versus our proposed prior (dashed blue line).}\label{fig:priorcomp_first}
\end{figure}

Figure \ref{fig:priorcomp_first} compares the proposed prior with the Scott \& Berger prior, in terms of prior mass versus model size, for the case of $d=30$ covariates. The uniform prior has been omitted from the plot as it would be represented by a horizontal line, reflecting a constant prior probability independent from the number of covariates included in the regression model. Whilst Scott \& Berger prior has a symmetrical behaviour, the proposed prior assigns more mass to the more simple models than to the more complex ones, as expected from expression \eqref{eq_ourprior}. \\

It is important to discuss some remarks about the proposed prior. First, the prior in \eqref{eq_ourprior} has practical relevance, especially in modern variable selection problems where the number of covariates can be significantly large. In fact, assigning less prior mass on large models reflects a parsimonious prior position, where the inclusion of additional covariates has to be supported by relatively strong evidence in the data.

The constant $c$ plays an important role. As $c\rightarrow0$, the prior \eqref{eq_ourprior} tends to be uniform, that is $p(M_{\pmb\gamma})\propto1$. If, on the other hand, we increase the value of $c$, the proposed prior will put higher mass on small models, yielding more parsimonious regression models. A natural choice is to set $c=1$, as if there is no prior information about the true model this appears to be the less informative choice. We empirically show in the simulation studies in Section \ref{sc_simulation} that $c=1$ leads better results, when compared to the uniform and the Scott \& Berger priors, if there is no prior information about the model size.

The proposed prior with the recommended value $c=1$ does not correct the multiplicity. One way of solving this issue, if required, would be to set constant $c$ as a function of the number of covariates. However, as mentioned in Section \ref{sc_intro}, we believe that for variable-selection problems the approach should be in line with the Bayesian framework of having prior and posterior probability representing, respectively, prior and posterior uncertainty. Although we agree that multiplicity correction is one way to define model prior probabilities in an objective sense, we argue that this is not a necessary condition for a model prior to satisfy, in particular, if the problem of interest is related to prediction.

As a final remark, we note that the prior based on loses induces a binomial prior for model size $k$:
\begin{eqnarray}\label{eq_uornormalised}
P(k) &=& \dbinom{d}{k}e^{-ck}\bigg/\sum_{l=0}^d\dbinom{d}{l}e^{-cl} \nonumber \\
&=& \dbinom{d}{k}\left(\frac{1}{e^c+1}\right)^k\left(\frac{e^c}{e^c+1}\right)^{d-k}.
\end{eqnarray}
Therefore, the marginal prior inclusion probability is given by
$$\omega = (e^c+1)^{-1}.$$
The immediate consequence of the above result is that the prior inclusion probability depends on the value of the constant $c$. If we set, as recommended for a minimally informative prior, $c=1$, then we obtain $\omega=0.27$. A smaller values of $c$ will lead a prior inclusion probability close to 1/2, that is $\omega\rightarrow1/2$ for $c\rightarrow0$. On the other hand, as $c$ increases, the prior inclusion probability becomes negligible and the model prior mass tends to be concentrated on the null model.

%--- SIMULATION STUDY ---------------------------------------------------------------------------
\section{Simulation study}\label{sc_simulation}
In this section we show the results of a simulation study designed to compare the performance of the proposed model prior based on losses with the uniform prior and the Scott \& Berger prior. It is important to highlight here that the value of $c$ considered for this simulation is one. The reason lies in the predictability of the behaviour of our prior for different values of $c$.

It is well known that a variable selection problem is driven by both the choice of the model prior and of the model-specific parameter prior. However, here the interest is in the effects on variable selection determined by the prior probability on the space of models. As such, the simulation exercise described has the purpose to analyse the frequentist properties of the posterior distribution on the model size and, to minimise any possible effects of the model-specific parameters prior, we choose to use the robust prior of \cite{Betal12} in conjunction with the three model priors. We obtained very similar results using the Zellner--Siow prior for the parameter of the regression (results are not included in this paper).

In line with objective Bayesian analysis, we have examined two key frequentist properties of the posterior distribution for the model size: the coverage and the mean squared error. The general idea is to perform repeated simulations from similar regression models. For the coverage, we compute the proportion of times the true model size is included in the 95\% credible interval of the posterior. We also compute the mean squared error from the mean and from the median of the true model size.

The simulation study encompasses a wide range of different scenarios, in terms of sample size, number covariates and model size, so to perform a thorough analysis. In detail, we have considered three sample sizes ($n=30$, $n=50$, and $n=100$) and four maximum number of covariates ($d=3$, $d=5$, $d=10$ and $d=15$) so to be able to compare the priors for different ratios number of covariates to sample size. To avoid cumbersome and difficult to interpret results, we have generated the repeated samples from models with, relatively to $d$, a small, medium and large number of covariates. This has been accomplished by randomly selecting the number of covariates according to a fixed prior inclusion probability $\omega$ of $0.15$, $0.50$ and $0.75$, respectively.

The above values of $n$, $d$ and $\omega$ give a total of 36 cases. For each case we repeat the following procedure for 100,000 times:
\begin{itemize}
\item Generate a design matrix $\mathbf{X}$ of size $d\times n$ where each element is an independent realisation of a standard normal distribution;
\item Generate a binary vector $\pmb\gamma$ from a sequence of $d$ independent Bernoulli experiments with probability of success equal to $\omega$;
\item From the robust prior in \eqref{eq_robustprior}, we generate the vector of coefficients $\pmb\beta_{\pmb\gamma}$;
\item We generate the response vector from the regression model in \eqref{eq_model}, considering $\phi=1$;
\item Finally, by using the above values of the design matrix and the vector of responses, we have computed the necessary quantities, including the marginal likelihoods, the model posteriors and the model size posterior distribution.
\end{itemize}
The last step of the above procedure has been performed under the three model priors considered in this paper: the uniform prior, the Scott \& Berger prior and the proposed prior.

With our simulation experience, the true model was equally identified under all the three considered priors. This part of the study has been performed by computing the average number of false positive and false negative in detecting the coefficients, both considering the highest probability model (HPM) and the median probability model (MPM). By false positive we consider a covariate which was included in any of the above models when it was not present in the true model. And by false negative we consider a covariate not in the true model but included in the MPM or the HPM. As the focus of this paper is on model prior and its performance on model size estimation the above results are not reported in detail. Instead, we discuss the effect of the model prior choice on the posterior model size distribution.

The detailed results of the simulations on the model size are presented in Table \ref{tab:simulation_summry}. Under each model prior, we report the coverage of the $95\%$ credible interval of the posterior, the mean squared error from the mean and the mean squared error from the median. From the experiments, the following aspects emerge.
The coverage under the Scott \& Berger prior appears to be the most stable across sample size and model size, although it is constantly above the nominal value of $95\%$. The uniform prior shows a coverage too low with respect to the nominal value when $d>5$ and $\omega=0.15$. This is more obvious for sample sizes of $n=30$ and $n=50$. The prior we propose appears to have a relatively low coverage when $\omega=0.75$. However, it shows an overall performance in the coverage closer to the nominal value than the Scott \& Berger's.

\begin{figure}[h]
\centering
\includegraphics[scale=0.8]{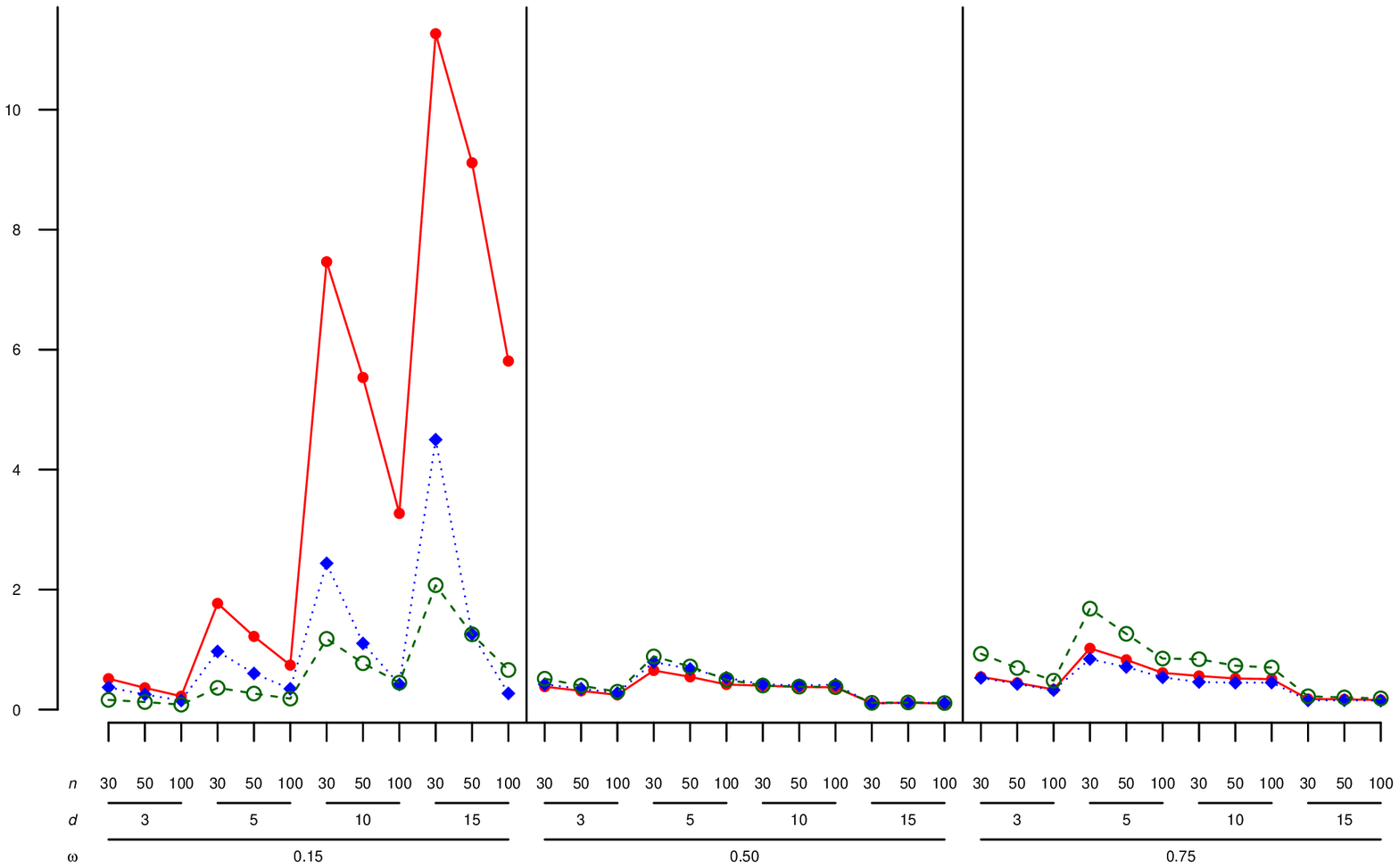}
\caption{Mean squared error from the mean from the simulation exercise on the posterior for the model size. The plot reports the results obtained by applying on the model space the uniform prior (red continuous line), Scott \& Berger prior (blue dotted line) and the proposed prior (dashed green line).}\label{fig_simulationsummary}
\end{figure}

To study the variability of posterior estimations, the mean squared errors from the posterior mean values and posterior median values are summarized in Table \ref{tab:simulation_summry}. As in terms of behaviour both mean squared errors have similar pattern, we examine in detail the mean squared error from the mean, which is also detailed in Figure \ref{fig_simulationsummary}. As one would expect, across all values of $d$ and $\omega$ the mean squared errors decrease as $n$ increases. If we focus on $\omega$, we note the following. For $\omega=0.15$, the uniform prior has the worse performance, while our proposed prior has generally the best performance. The only exception being the scenario with $n=100$ and $d=15$. The three priors have very similar performance for expected medium size regression models, i.e. $\omega=0.50$. Finally, for relatively large models, our prior is outperformed by both the uniform prior and the Scott \& Berger prior. However, in this latter case, the differences are not as prominent as the case of relatively small models and tend to disappear for large values of $d$.
\begin{table}[h!]
\centering
\begin{adjustbox}{max width=\textwidth}
\begin{tabular}{lcccccccccc}
\hline
                      $\mathbf{n=30}$ & \multicolumn{3}{c}{\textbf{Uniform}}      & \multicolumn{3}{c}{\textbf{Scott \& Berger}} & \multicolumn{3}{c}{\textbf{Proposed, $c = 1$}}   &        \\
                      $d$ & Coverage & MSE Mean & MSE Median & Coverage  & MSE Mean  & MSE Median  & Coverage & MSE Mean & MSE Median & $\omega$ \\
                       \hline
\multicolumn{1}{c}{3}  & 0.995    & 0.515    & 0.445      & 0.998     & 0.371     & 0.297       & 0.999    & 0.162    & 0.163      & 0.15   \\
                       & 0.995    & 0.382    & 0.507      & 0.998     & 0.421     & 0.605       & 0.967    & 0.515    & 0.685      & 0.50   \\
                       & 0.989    & 0.545    & 0.746      & 0.997     & 0.529     & 0.802       & 0.908    & 0.930    & 1.166      & 0.75   \\
\multicolumn{1}{c}{5}  & 0.980    & 1.770    & 1.640      & 0.995     & 0.969     & 0.784       & 0.997    & 0.363    & 0.323      & 0.15   \\
                       & 0.992    & 0.652    & 0.763      & 0.997     & 0.793     & 1.059       & 0.950    & 0.886    & 1.163      & 0.50   \\
                       & 0.960    & 1.021    & 1.221      & 0.995     & 0.843     & 1.163       & 0.852    & 1.682    & 2.014      & 0.75   \\
\multicolumn{1}{c}{10} & 0.773    & 7.464    & 7.444      & 0.993     & 2.438     & 1.760       & 0.989    & 1.179    & 1.011      & 0.15   \\
                       & 0.990    & 0.398    & 0.475      & 0.991     & 0.421     & 0.480       & 0.966    & 0.401    & 0.493      & 0.50   \\
                       & 0.957    & 0.561    & 0.706      & 0.990     & 0.460     & 0.593       & 0.886    & 0.840    & 0.976      & 0.75   \\
\multicolumn{1}{c}{15} & 0.786    & 11.266   & 11.314     & 0.992     & 4.501     & 3.656       & 0.973    & 2.074    & 1.926      & 0.15   \\
                       & 0.996    & 0.107    & 0.129      & 0.996     & 0.102     & 0.117       & 0.974    & 0.113    & 0.130      & 0.50   \\
                       & 0.972    & 0.177    & 0.212      & 0.993     & 0.153     & 0.185       & 0.918    & 0.222    & 0.249      & 0.75   \\
\hline
\multicolumn{4}{c}{\vspace{0.1cm}}\\
\hline
                     $\mathbf{n=50}$  & \multicolumn{3}{c}{\textbf{Uniform}}      & \multicolumn{3}{c}{\textbf{Scott \& Berger}} & \multicolumn{3}{c}{\textbf{Proposed, $c = 1$}}   &        \\
					  $d$ & Coverage & MSE Mean & MSE Median & Coverage  & MSE Mean  & MSE Median  & Coverage & MSE Mean & MSE Median & $\omega$ \\
					  \hline
\multicolumn{1}{c}{3}  & 0.997    & 0.363    & 0.298      & 0.999     & 0.252     & 0.206       & 0.999    & 0.126    & 0.134      & 0.15     \\
                       & 0.995    & 0.312    & 0.417      & 0.998     & 0.347     & 0.478       & 0.971    & 0.399    & 0.524      & 0.50     \\
                       & 0.988    & 0.445    & 0.618      & 0.995     & 0.427     & 0.635       & 0.932    & 0.693    & 0.867      & 0.75     \\
\multicolumn{1}{c}{5}  & 0.988    & 1.220    & 1.131      & 0.997     & 0.603     & 0.478       & 0.998    & 0.267    & 0.248      & 0.15     \\
                       & 0.992    & 0.546    & 0.665      & 0.995     & 0.678     & 0.888       & 0.954    & 0.719    & 0.956      & 0.50     \\
                       & 0.963    & 0.830    & 1.045      & 0.993     & 0.710     & 0.989       & 0.878    & 1.263    & 1.548      & 0.75     \\
\multicolumn{1}{c}{10} & 0.882    & 5.536    & 5.382      & 0.997     & 1.104     & 0.746       & 0.996    & 0.773    & 0.629      & 0.15     \\
                       & 0.991    & 0.378    & 0.456      & 0.993     & 0.404     & 0.463       & 0.970    & 0.383    & 0.482      & 0.50     \\
                       & 0.960    & 0.519    & 0.659      & 0.990     & 0.448     & 0.581       & 0.896    & 0.732    & 0.856      & 0.75     \\
\multicolumn{1}{c}{15} & 0.784    & 9.115    & 9.133      & 0.996     & 1.259     & 0.808       & 0.990    & 1.255    & 1.076      & 0.15     \\
                       & 0.995    & 0.115    & 0.133      & 0.996     & 0.113     & 0.128       & 0.974    & 0.119    & 0.138      & 0.50     \\
                       & 0.975    & 0.171    & 0.201      & 0.994     & 0.156     & 0.188       & 0.918    & 0.203    & 0.231      & 0.75 \\
\hline
\multicolumn{4}{c}{\vspace{0.1cm}}\\
\hline
					$\mathbf{n=100}$ & \multicolumn{3}{c}{\textbf{Uniform}}      & \multicolumn{3}{c}{\textbf{Scott \& Berger}} & \multicolumn{3}{c}{\textbf{Proposed, $c = 1$}}   &        \\
                     $d$ & Coverage & MSE Mean & MSE Median & Coverage  & MSE Mean  & MSE Median  & Coverage & MSE Mean & MSE Median & $\omega$ \\
                     \hline
\multicolumn{1}{c}{3}  & 0.999    & 0.223    & 0.174   & 0.999      & 0.151      & 0.120     & 0.999     & 0.083    & 0.084   & 0.15     \\
                       & 0.995    & 0.245    & 0.323   & 0.995      & 0.277      & 0.364     & 0.978     & 0.298    & 0.382   & 0.50     \\
                       & 0.987    & 0.336    & 0.461   & 0.993      & 0.325      & 0.471     & 0.949     & 0.482    & 0.596   & 0.75     \\
\multicolumn{1}{c}{5}  & 0.990    & 0.741    & 0.619   & 0.997      & 0.347      & 0.269     & 0.998     & 0.183    & 0.170   & 0.15     \\
                       & 0.993    & 0.418    & 0.532   & 0.992      & 0.531      & 0.661     & 0.967     & 0.499    & 0.658   & 0.50     \\
                       & 0.967    & 0.613    & 0.810   & 0.994      & 0.537      & 0.764     & 0.909     & 0.851    & 1.051   & 0.75     \\
\multicolumn{1}{c}{10} & 0.954    & 3.270    & 2.989   & 0.998      & 0.420      & 0.279     & 0.997     & 0.449    & 0.340   & 0.15     \\
                       & 0.991    & 0.372    & 0.441   & 0.993      & 0.415      & 0.476     & 0.970     & 0.375    & 0.471   & 0.50     \\
                       & 0.958    & 0.508    & 0.647   & 0.990      & 0.451      & 0.580     & 0.894     & 0.701    & 0.821   & 0.75     \\
\multicolumn{1}{c}{15} & 0.866    & 5.810    & 5.716   & 0.997      & 0.269      & 0.152     & 0.994     & 0.660    & 0.530   & 0.15     \\
                       & 0.996    & 0.102    & 0.120   & 0.996      & 0.108      & 0.123     & 0.977     & 0.109    & 0.125   & 0.50     \\
                       & 0.975    & 0.160    & 0.187   & 0.994      & 0.149      & 0.178     & 0.922     & 0.187    & 0.210   & 0.75 \\
\hline
\end{tabular}
\end{adjustbox}
\caption{Summary of simulation results for the three priors: uniform prior, Scott \& Berger prior and the proposed prior with $c=1$. A proportion of identifying the true number of covariates within the $95\%$ of the posterior probability (Coverage) and the mean squared error of posterior mean values (MSE Mean) and posterior median values (MSE Median) are computed.}
\label{tab:simulation_summry}
\end{table}

Overall, it appears that the proposed prior with $c=1$ has the most stable frequentist performance, making it appealing in scenarios of minimal prior knowledge about the true model size. On the other hand, if one had prior information about the true models size being close to the full model, a value of $c$ smaller than one could be selected (e.g. $c=0.5$) so to obtain a model prior more similar to the uniform prior and increase performance.

\clearpage

%--- REAL DATA ----------------------------------------------------------------------------------
\section{Illustrative examples with real data sets}\label{sc_realdata}

In this section we investigate the properties of objective model priors for variable selection in real data sets. The considered data sets are US crime data \citep{V78} and the Hald data \citep{Woods32}, which have been extensively used in literature (see \cite{Kubi96} and \cite{Letal08}, for example). At last, we perform a robustness analysis to highlight the sensitivity of the inferential results to relatively small changes to the data.

\subsection{US crime data}\label{sc_uscrime}

The first data set consists of $n = 47$ observations for $d = 15$ covariates related to crime data in the US for 1960 \citep{V78}. The aim of the analysis is to understand which variables, such as population type, police expenditure and ethnicity (among others), impact on the crime rate per head in the population (the response variable).

\begin{figure}[h!]
\centering
\subfigure[]{%
\includegraphics[scale=0.40]{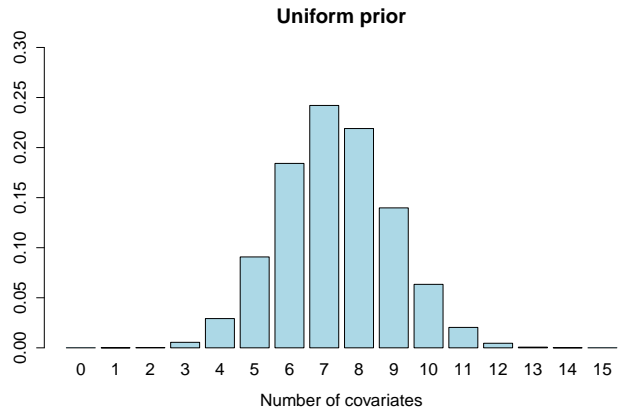}  
\label{fig:subfigur12}}
\subfigure[]{%
\includegraphics[scale=0.40]{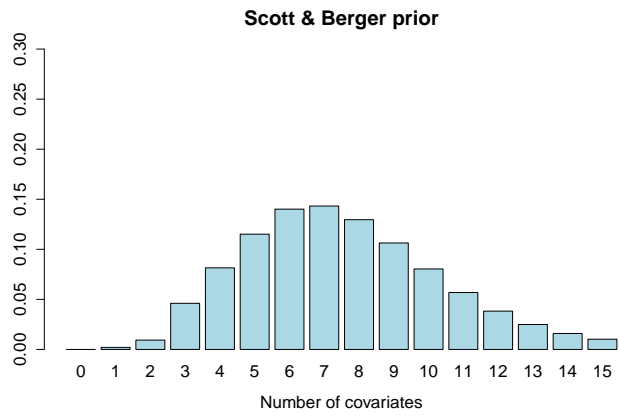} 
\label{fig:subfigur22}}
\subfigure[]{%
\includegraphics[scale=0.40]{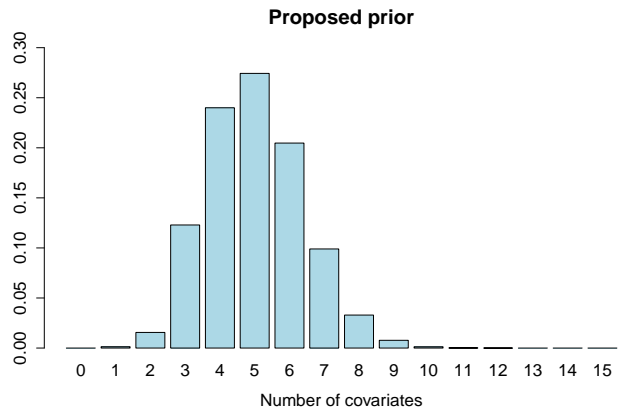}  
\label{fig:subfigur1}
}
\caption{Posterior distribution of the number of covariates for the US Crime data set. Three model priors are considered: uniform, Scott \& Berger and proposed prior with $c=1$.} \label{fig:US_post}
\end{figure}

The summary statistics of the model size posterior distributions are presented in Table \ref{tab:US1}, where we considered the proposed prior with different values of the constant $c$. The plot of the posterior distributions for the model size are represented in Figure \ref{fig:US_post}. The mean and the median obtained by the use of the uniform prior and the Scott \& Berger prior are similar. The proposed prior, with $c=1$, gives mean and median model size smaller than the ones obtained by using the other two priors. We note that the posterior mean and median model size increase when $c$ is smaller than one, as expected, and for $c>1$ these values decrease as the prior puts more mass on smaller models. Again, this result is in line on what discussed in Sections \ref{sc_priors} and \ref{sc_simulation}. From the table, and by inspecting the the posterior plots, we see that the Scott \& Berger prior yields a posterior for the model size with higher uncertainty than the one obtained by using either the uniform prior or the proposed prior.

The HPM identified by the uniform prior includes 6 covariates, unlike the one determined by Scott \& Berger prior and the proposed prior, which both assign the highest mass to a model of size 3. The posterior probability for the HPM is of 0.019 for the uniform prior, 0.021 for the Scott \& Berger prior, and 0.056 when the proposed prior (with $c=1$) is used. In Table \ref{tab:US_postinclu} we can see what covariates are included in the HPM under each one of the considered priors, and that both Scott \& Berger and our prior agree in the result. In Table \ref{tab:US_postinclu}, the proposed model prior tends to assign lower probabilities compared to the other two priors, particularly for covariates which are unlikely to be included (posterior inclusion probabilities less than 1/2). The main result is that the covariates concerning age of males, unemployment rate among male and probability of imprisonment are included in the MPM under the uniform and the Scott \& Berger priors, but not under the proposed prior.

Although the proposed prior and Scott \& Berger's disagree on the size of the MPM, we note that both values, 3 and 6, are well within the 95\% credible interval of the respective posteriors. The same can not be said for the uniform prior, as the corresponding credible interval does not contain the size 3.

\begin{table}[h!]
\centering
\begin{tabular}{lcccc| c c}
\hline 
Model prior & Mean & Median & SD & $95\%$ C.I. & HPM & MPM \\ 
\hline 
Uniform & 7.32 & 7 & 1.78 & (4,11) & 6 & 6\\ 
%\hline 
Scott \& Berger & 7.49 & 7 & 2.91 & (3,14) & 3 & 6 \\ 
%\hline 
Proposal($c=1.0$) & 5.00 & 5 & 1.56 & (3,8) & 3 & 3\\ 
%\hline 
Proposal($c=0.5$) & 6.06 & 6 & 1.72 & (4,10) & 3  & 3 \\ 
%\hline 
Proposal($c=1.5$) & 4.13 & 4 & 1.42 & (2,7) & 3 & 3 \\ 
%\hline 
Proposal($c=2.0$) & 3.45 & 3 & 1.29 & (1,6) & 3 & 3 \\ 
\hline 
\end{tabular}
\caption{Comparison of the posterior summary statistics for the US Crime data set. Four statistics for the number of covariates are measured: mean, median, standard deviation (SD) and the $95\%$ confidence interval ($95\%$ C.I.). The number of covariates included in the highest posterior probability model (HPM) and the median probability model (MPM) are reported. }
\label{tab:US1}
\end{table}

\begin{table}
\centering
\begin{tabular}{lccc}
\hline 
 & \multicolumn{3}{c}{Posterior Inclusion Probability} \\ 
%\hline 
Covariate & Uniform & Scott \& Berger & Proposed ($c=1$) \\ 
\hline 
Percentage of males aged 14-24 & \textbf{0.76}$ \bullet$ & \textbf{0.70}$\:\:\:$ & 0.46$\:\:\:$ \\ 
%\hline 
Indicator variable for a Southern state & 0.25$\:\:\:$ & 0.28$\:\:\:$ & 0.10$\:\:\:$ \\ 
%\hline 
Mean years of schooling & \textbf{0.88}$ \bullet$ & \textbf{0.85}$ \bullet$ & \textbf{0.72}$ \bullet$ \\ 
%\hline 
Police expenditure in 1960 & \textbf{0.83}$ \bullet$ & \textbf{0.84}$ \bullet$ & \textbf{0.81}$ \bullet$ \\ 
%\hline 
Police expenditure in 1959 & 0.38$\:\:\:$ & 0.41$\:\:\:$ & 0.27$\:\:\:$ \\ 
%\hline 
Labour force participation rate & 0.23$\:\:\:$ & 0.27$\:\:\:$ & 0.11$\:\:\:$ \\ 
%\hline 
Number of males per 1000 females & 0.38$\:\:\:$ & 0.42$\:\:\:$ & 0.29$\:\:\:$ \\ 
%\hline 
State population & 0.28$\:\:\:$ & 0.31$\:\:\:$ & 0.12$\:\:\:$ \\ 
%\hline 
Number of non-whites per 1000 people & 0.23$\:\:\:$ & 0.27$\:\:\:$ & 0.10$\:\:\:$ \\ 
%\hline 
Unemployment rate of urban males 14-24 & 0.30$\:\:\:$ & 0.34$\:\:\:$ & 0.11$\:\:\:$ \\ 
%\hline 
Unemployment rate of urban males 35-39 & \textbf{0.52}$ \bullet$ & \textbf{0.51}$\:\:\:$ & 0.21$\:\:\:$ \\ 
%\hline 
Gross domestic product per head & 0.36$\:\:\:$ & 0.37$\:\:\:$ & 0.18$\:\:\:$ \\ 
%\hline 
Income inequality & \textbf{0.98}$ \bullet$ & \textbf{0.97}$ \bullet$ & \textbf{0.96}$ \bullet$ \\ 
%\hline 
Probability of imprisonment & \textbf{0.70}$ \bullet$ & \textbf{0.66}$\:\:\:$ & 0.44$\:\:\:$ \\ 
%\hline 
Average time served in state prisons & 0.25$\:\:\:$ & 0.28$\:\:\:$ & 0.12$\:\:\:$ \\ 
\hline 
\end{tabular}
\caption{Posterior inclusion probabilities for the US Crime data set. The covariates with posterior inclusion probabilities greater than 1/2 are highlighted in bold, and a dot notation represents the covariate included in the highest posterior probability model.}
\label{tab:US_postinclu} 
\end{table}

\clearpage

\clearpage

\subsection{Hald data}\label{sc_hald}
The Hald data set contains $n=13$ observations with $d=4$ covariates, and it concerns an engineering application to study the cement composition \citep{Woods32}. In particular, the study considers the effect on the heat evolved per gram of cement (in calories) by the amount of tricalcium aluminate, the amount of tricalcium silicate, the amount of tricalcium aluminio ferrite and the amount of dicalcium silicate.

\begin{figure}[h!]
\centering
\subfigure[]{%
\includegraphics[scale=0.22]{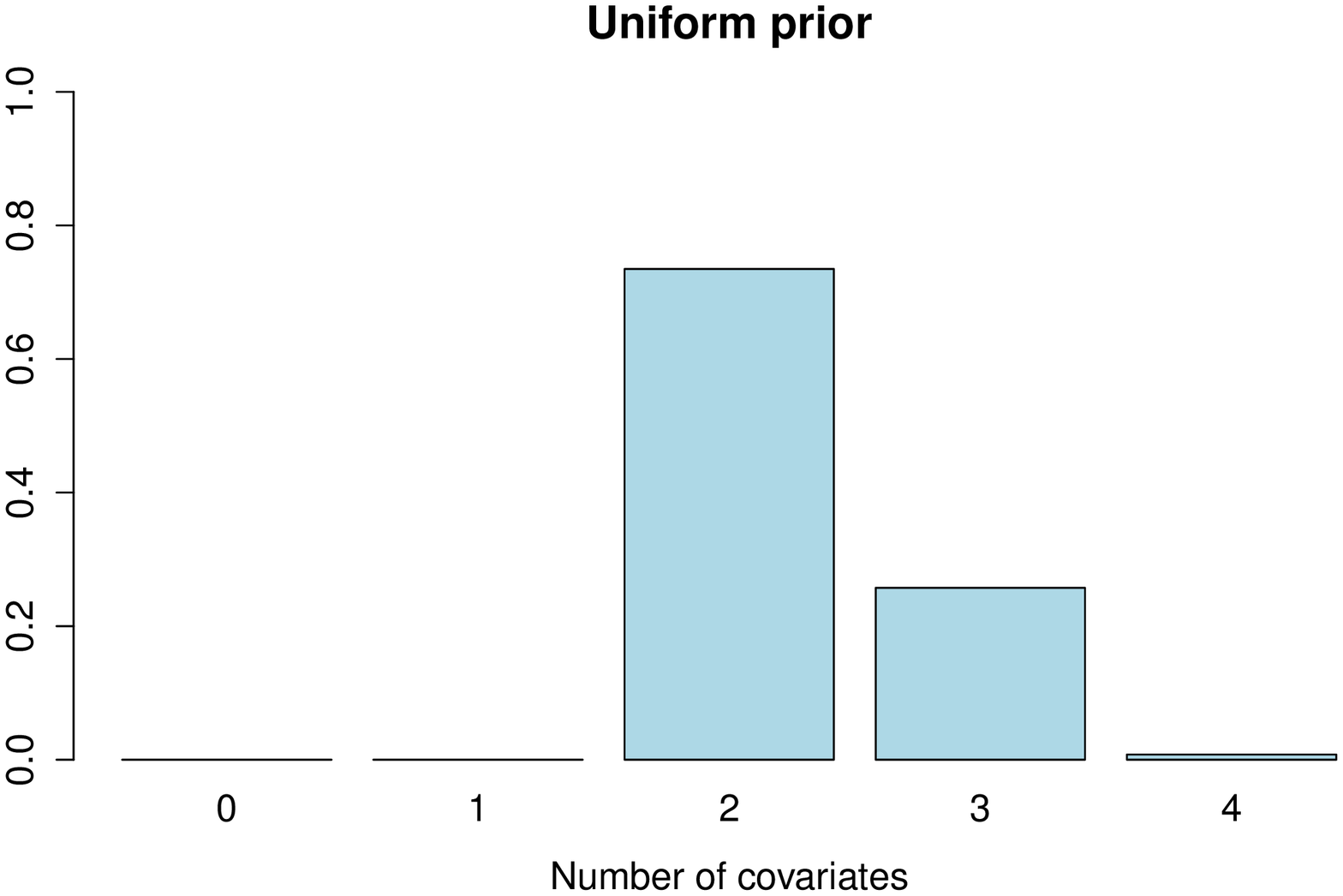}  
\label{fig:hald_sub12}}
\subfigure[]{%
\includegraphics[scale=0.22]{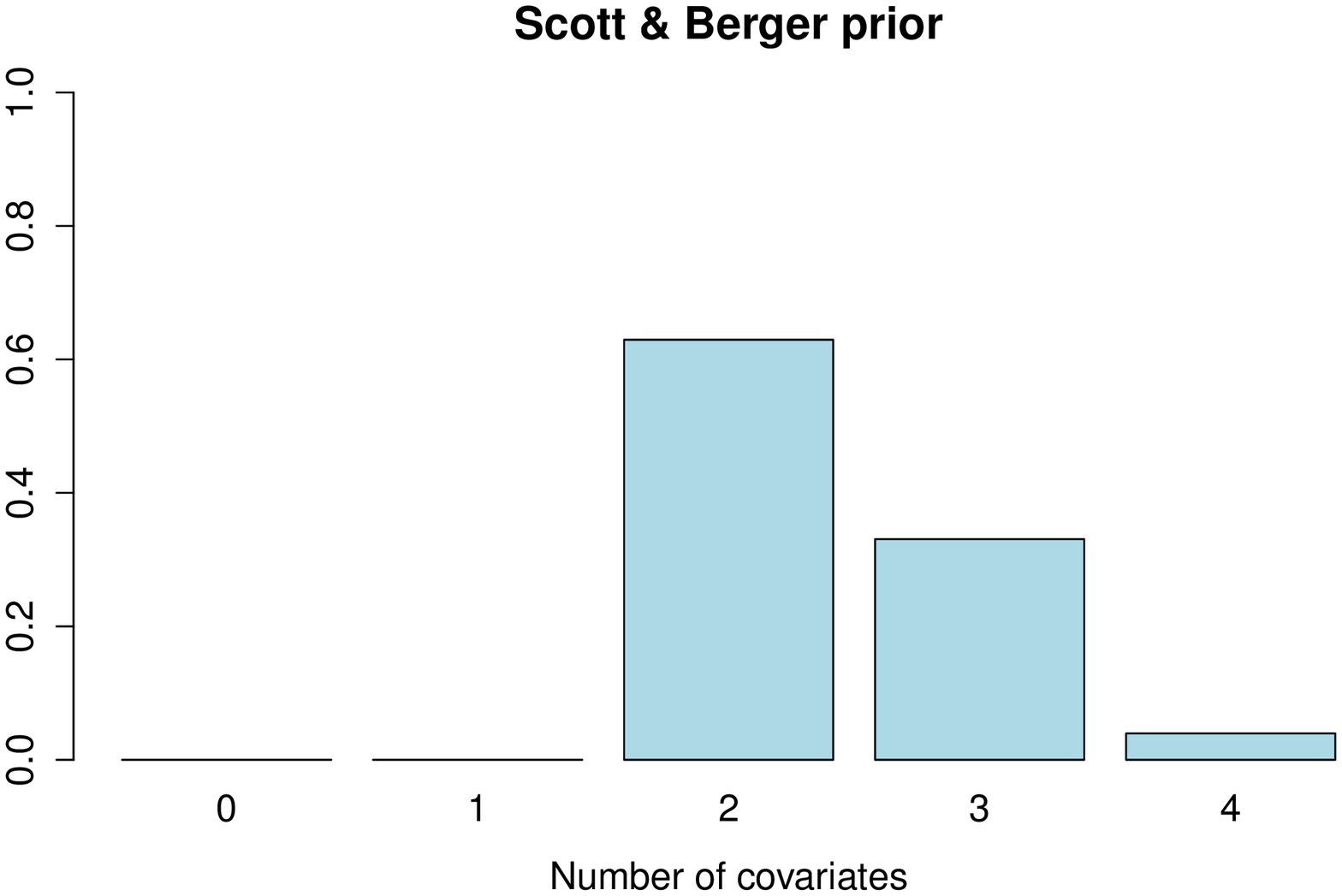} 
\label{fig:hald_sub2}}
\subfigure[]{%
\includegraphics[scale=0.22]{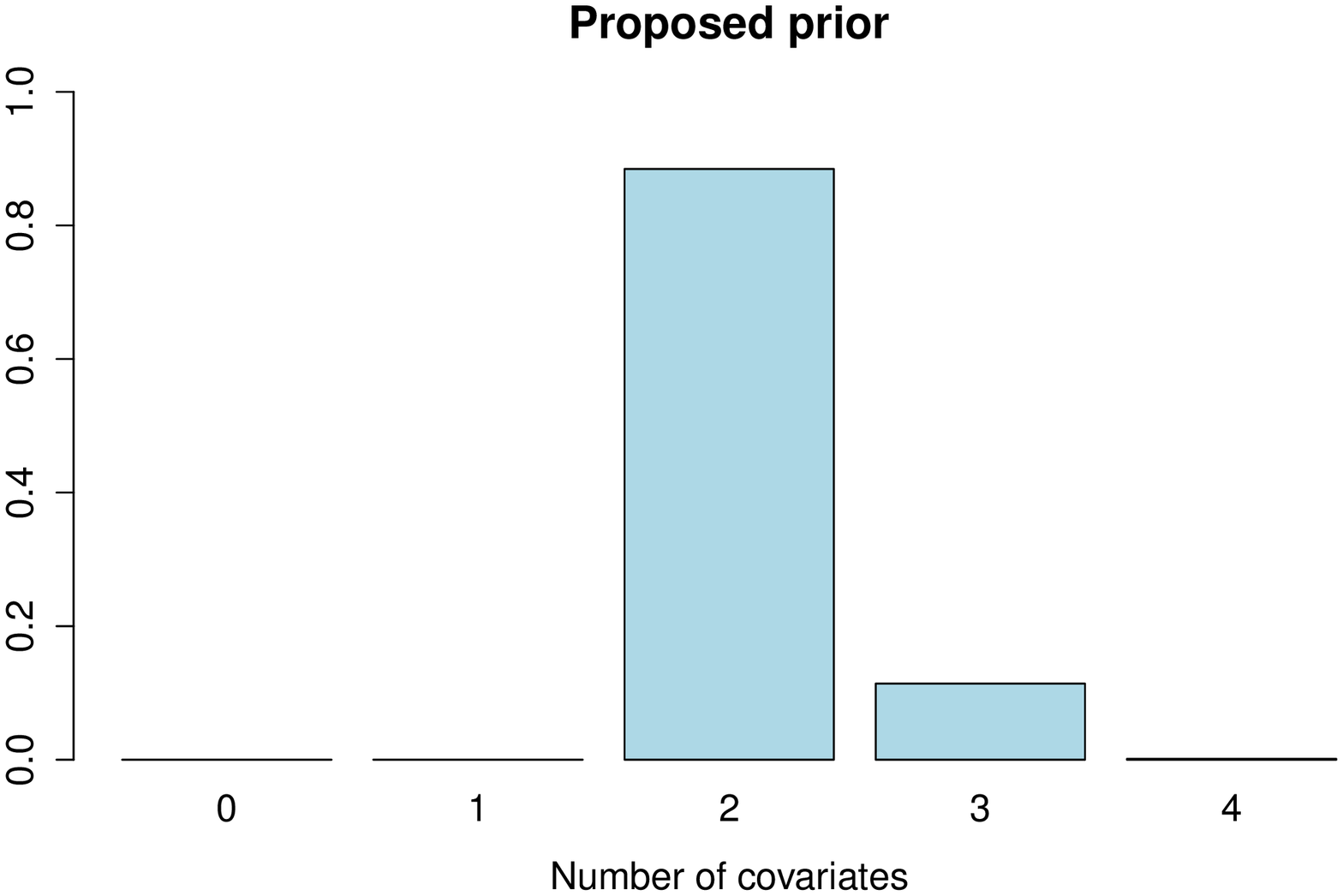}  
\label{fig:hald_sub1}}
\caption{Posterior distribution of the number of covariates for the Hald data set. Three model priors are considered: uniform, Scott \& Berger and proposed prior with $c=1$.}
\label{fig:hald_post}
\end{figure}

The variable selection problem for the Hald data turns out to be a much simpler task than the one discussed in Section \ref{sc_uscrime}. In fact, by examining both Table \ref{tab:hald_summary} and Figure \ref{fig:hald_post}, we note that the size of HPM and MPM are 2 under each considered prior. The only exception being the proposed prior with $c=2$. In this case, as expected, the prior is extremely parsimonious assigning high mass on the null model, in comparison to the mass assigned to any other regression model. There is a slightly higher variability in the posterior of the model size under the Scott \& Berger prior, and this is reflected by both a higher standard deviation and a larger credible interval. The coinciding result on the model size estimation is supported by the similar posterior inclusion probabilities in Table \ref{tab:hald_postinclu}. The posterior inclusion probability assigned to the first two covariates is definitely above 1/2 under each prior, whilst the one assigned to the remaining two covariates is clearly below 1/2.

The Hald data set appears to represent a case where the information contained in the data about which covariate should be included is so strong that the choice of the model prior does not make much difference; provided, of course, that this choice is not too extreme. The above is reflected in the posterior probability associated to the HPM being close or above 50\% under each prior. In fact, the probability is 0.55 under the uniform prior, 0.47 under the Scott \& Berger prior and 0.67 under the proposed prior.
\begin{table}[h!]
\centering
\begin{tabular}{ccccc|cc}
\hline 
 & Mean & Median & SD & $95\%$ C.I. & HPM & MPM \\ 
\hline 
Uniform & 2.27 & 2 & 0.74 & (2,3) & 2 & 2 \\ 
%\hline 
Scott \& Berger & 2.41 & 2 & 0.81 & (2,4) & 2 & 2 \\ 
%\hline 
Proposal ($c=1$) & 2.12 & 2 & 0.72 & (2,3) & 2 & 2 \\ 
%\hline 
Proposal ($c=0.5$) & 2.25 & 2 & 0.73 & (2,3) & 2 & 2 \\ 
%\hline  
Proposal ($c=1.5$) & 2.07 & 2 & 0.71 & (2,3) & 2 & 2\\ 
%\hline 
Proposal ($c=2$) & 0.02 & 0 & 1.15 & (0,0) & 2 & 2 \\ 
\hline 
\end{tabular} 
\caption{Comparison of the posterior summary statistics for the Hald data set. Four statistics for the number covariates are measured: mean, median, standard deviation (SD) and the $95\%$ confidence interval ($95\%$ C.I.). The number of covariates included in the highest posterior probability model (HPM) median probability model (MPM) are reported.}
\label{tab:hald_summary}
\end{table}

\begin{table}
\centering
\begin{tabular}{lccc}
\hline 
 & \multicolumn{3}{c}{Posterior Inclusion Probability} \\ 
%\hline 
Covariate & Uniform & Scott \& Berger & Proposed ($c=1$) \\ 
\hline 
Ticalcium aluminate & \textbf{0.98}$ \bullet$ & \textbf{0.98}$ \bullet$ & \textbf{0.99}$ \bullet$ \\ 
%\hline 
Tricalcium silicate & \textbf{0.75}$ \bullet$ & \textbf{0.76}$ \bullet$ & \textbf{0.75}$ \bullet$ \\ 
%\hline 
Tetracalcium alumino ferrite & 0.18$\:\:\:$ & 0.26$\:\:\:$ & 0.08$\:\:\:$ \\ 
%\hline 
Dicalcium silicate & 0.36$\:\:\:$ & 0.42$\:\:\:$ & 0.30$\:\:\:$ \\ 
\hline 
\end{tabular}
\caption{Posterior inclusion probabilities for the Hald data set. The covariates with a posterior inclusion probability greater than 1/2 are highlighted in bold, and a dot notation represents the covariate included in the highest posterior probability model.}
\label{tab:hald_postinclu}
\end{table}

\subsection{Robustness of model priors}
To analyse the robustness of the proposed prior to small changes in the data, and to compare the results with the uniform and the Scott \& Berger priors, we look at the performance in estimating the model size and the posterior inclusion probabilities for repeated random sub-samples of the original data, both for the US Crime data and the Hald data.

The general idea is to extract 500 sub-samples of a size that is the 85\% of the numerosity of the data set, and perform variable selection as done in Section \ref{sc_realdata} using these sub-samples as observations. Then we compare the performance of the model priors by considering the histogram of the posterior mean model size, and the box-plots of the posterior inclusion probabilities for the considered priors.

\begin{figure}[h!]
\centering
\subfigure[]{%
\includegraphics[scale=0.30]{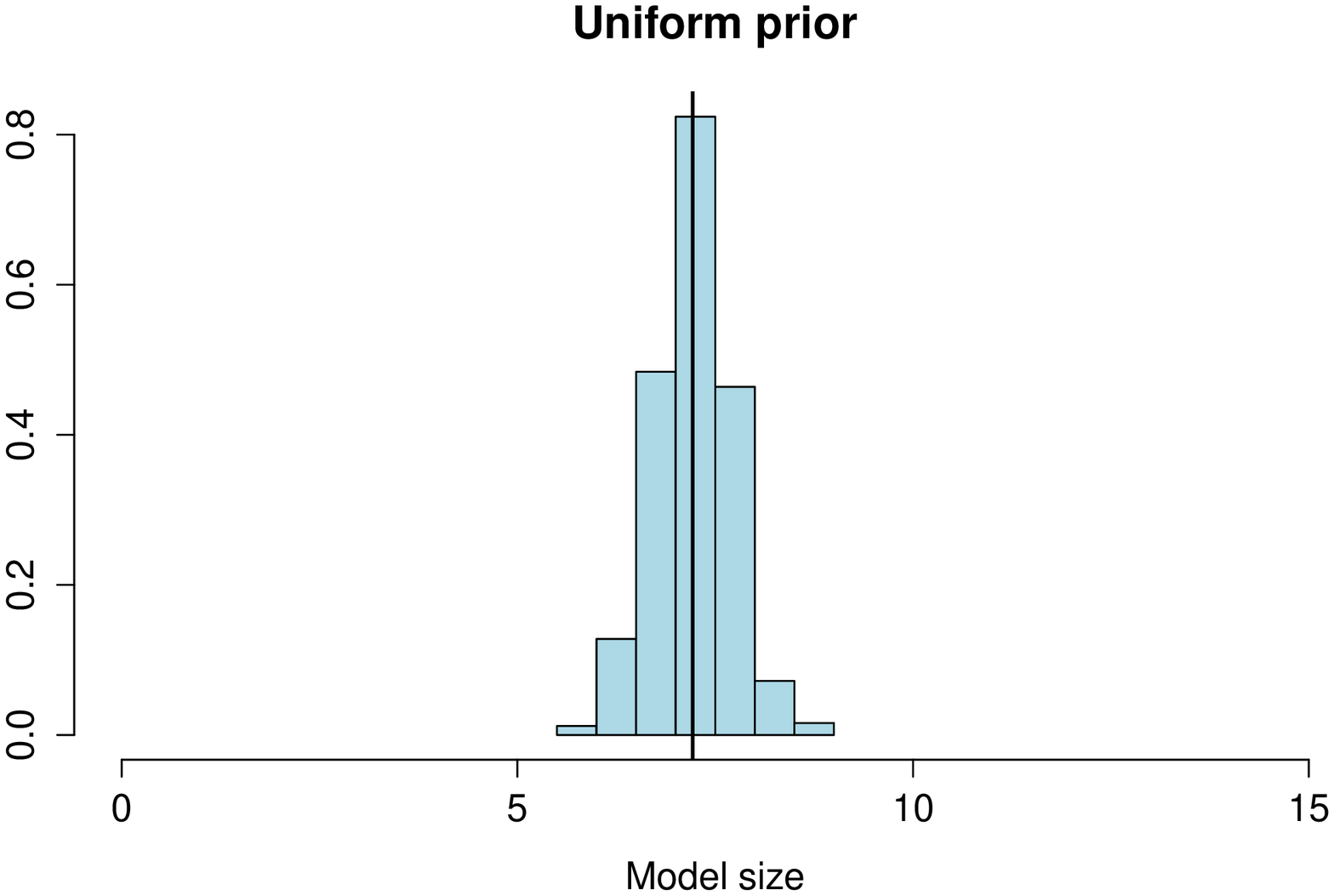}  
\label{US_RA_hists:subfigur1}}
\quad
\subfigure[]{%
\includegraphics[scale=0.30]{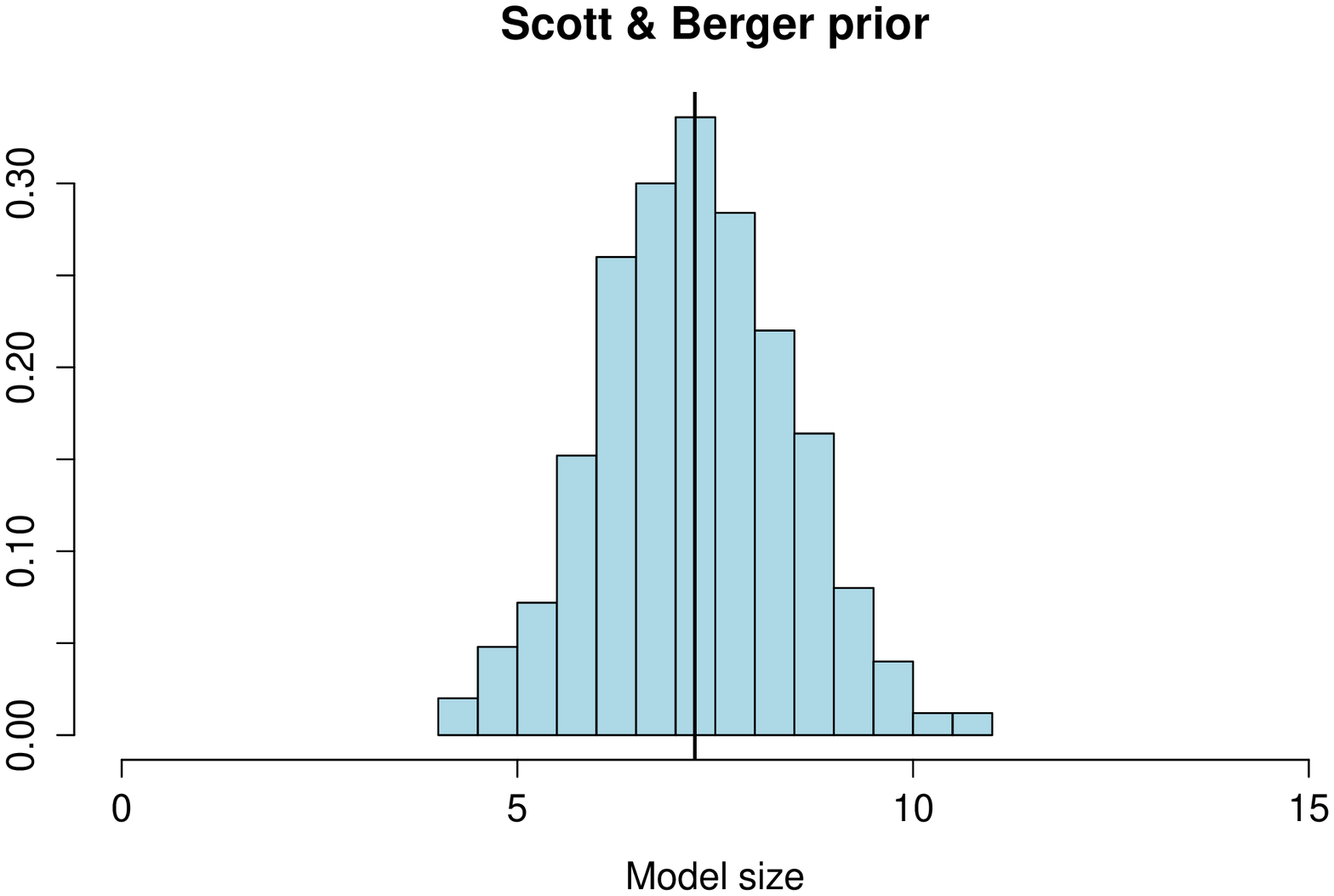} 
\label{US_RA_hists:subfigur2}}
\subfigure[]{%
\includegraphics[scale=0.30]{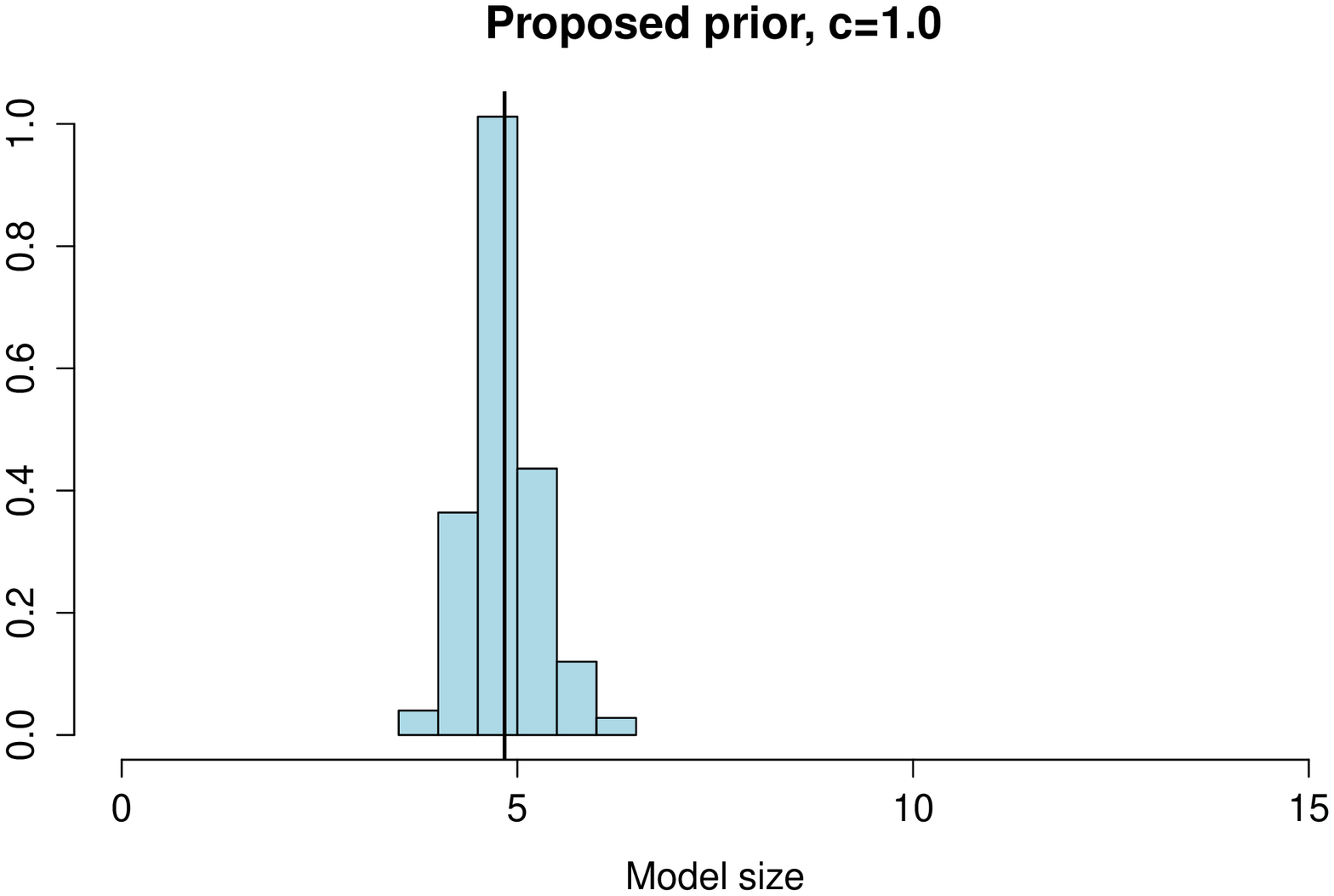}  
\label{US_RA_hists:subfigur3}}
\quad
\subfigure[]{%
\includegraphics[scale=0.30]{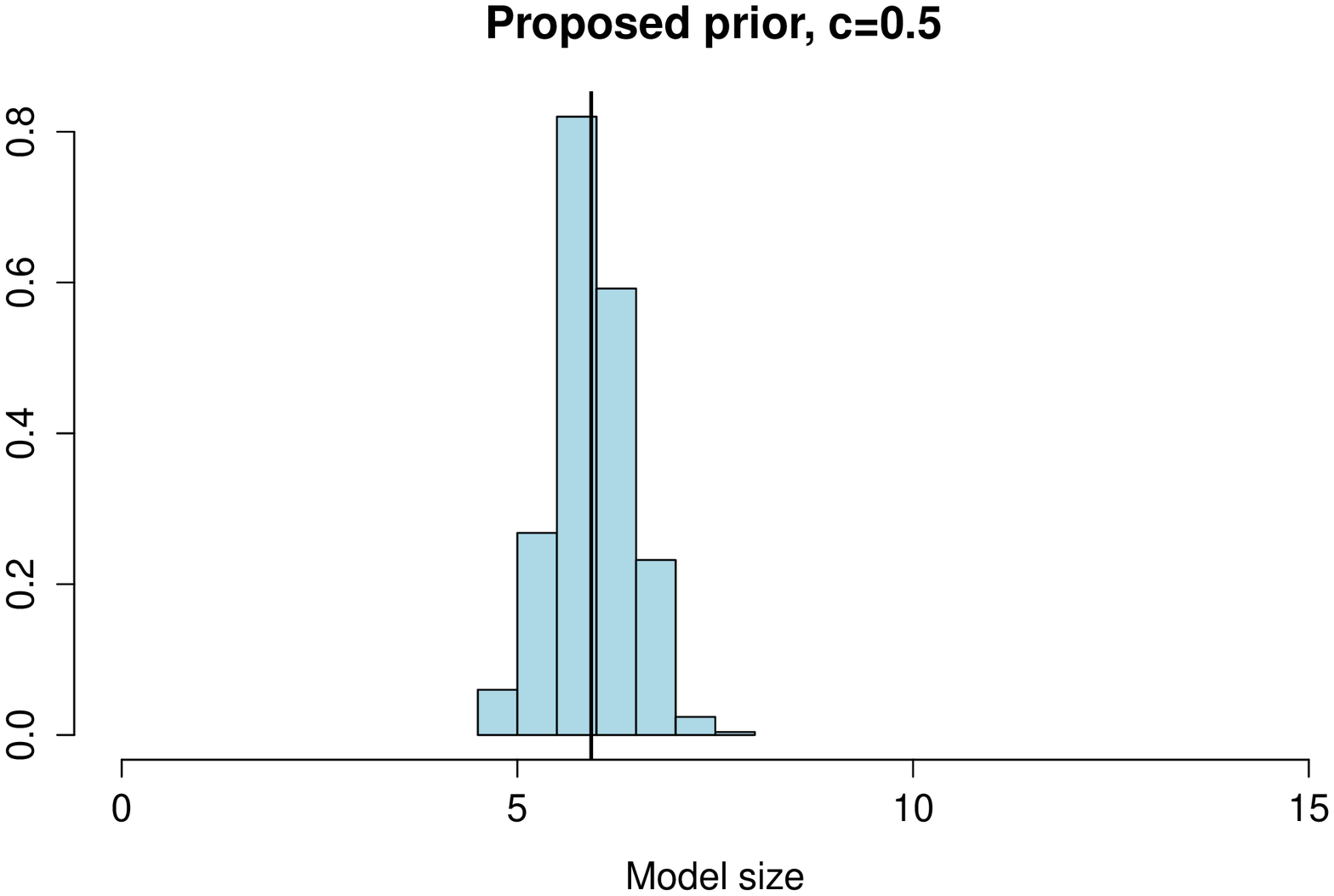} 
\label{US_RA_hists:subfigur4}}
\subfigure[]{%
\includegraphics[scale=0.30]{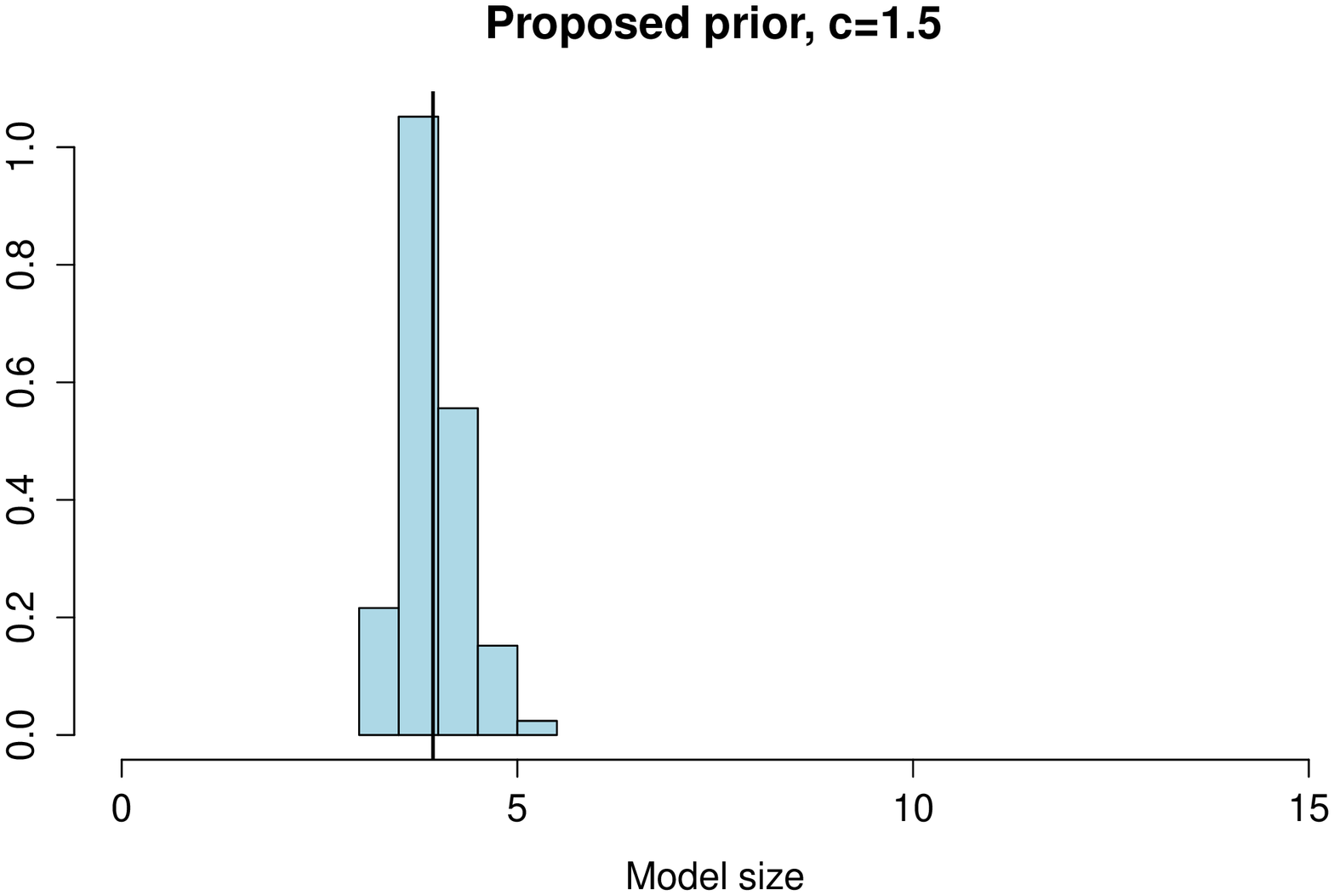}  
\label{US_RA_hists:subfigur5}}
\quad
\subfigure[]{%
\includegraphics[scale=0.30]{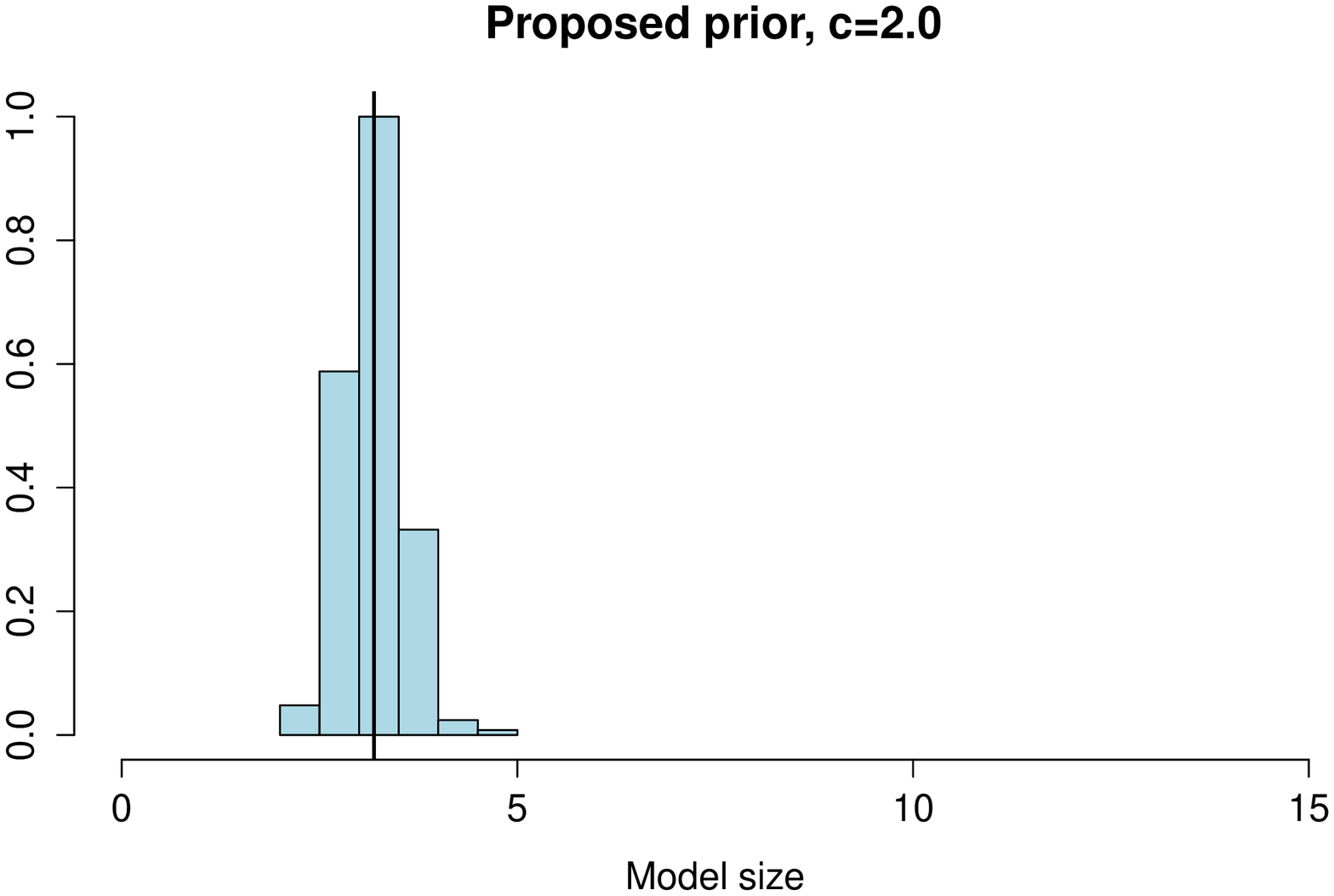}  
\label{US_RA_hists:subfigur6}}
\caption{Histograms of the posterior mean model size for the considered priors, where the proposed prior has been analysed for $c = 1.0$, $c = 0.5$, $c = 1.5$ and $c = 2.0$ (500 sub-samples of size 40 from the US Crime data set).}
\label{fig:US_RA_hists}
\end{figure}

For the US Crime data set analysed in Section \ref{sc_uscrime}, we consider 500 samples of size 40. The analysis takes also into consideration the results for different values of $c$ for the proposed prior. Figure \ref{fig:US_RA_hists} shows the histogram of the mean model size when we apply the uniform prior and the Scott \& Berger; for the proposed prior, we consider $c = 1.0$, $c = 0.5$, $c = 1.5$ and $c = 2.0$. In the plots we have indicated with a vertical line the estimated model size when the whole data set is considered. We note the following. First, the uncertainty in the model size is larger when the Scott \& Berger prior is used. All the remaining cases appear to lead similar variability in the posterior model size estimates. Second, there is a sensitivity of our prior to $c$, with an impact on the mean location of the distribution, in the sense that when $c$ increases the mean model size decreases. This agrees with what discussed in Section \ref{sc_priors} about the parsimony of the prior when $c$ increases. Also, note that proposed prior tends to yield a posterior similar to the posterior yielded by the uniform prior for $c=0.5$, as expected.

\begin{figure}[h!]
\centering
\subfigure[]{%
\includegraphics[scale=0.30]{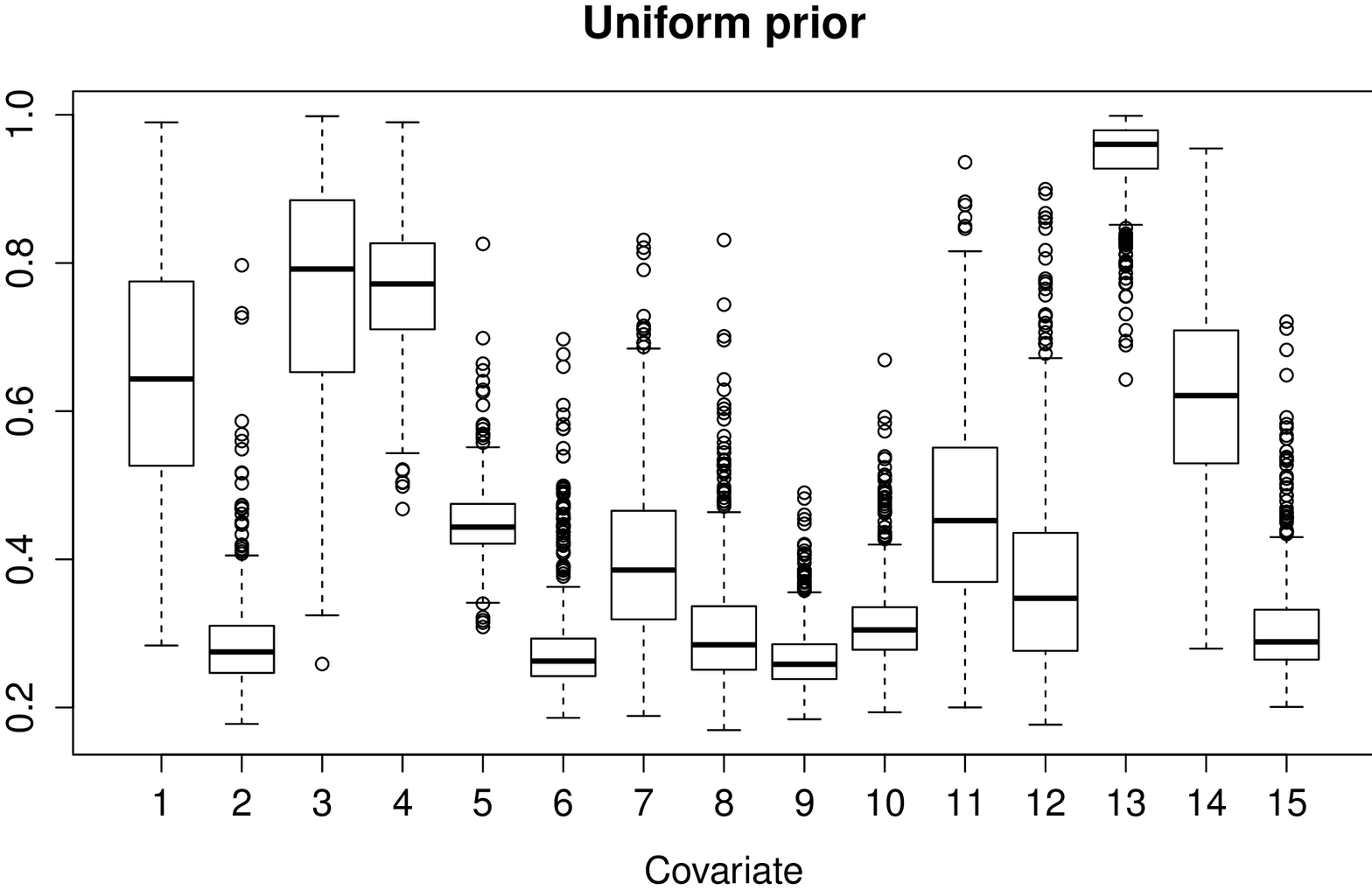}  
\label{US_RA_box:subfigur1}}
\quad
\subfigure[]{%
\includegraphics[scale=0.30]{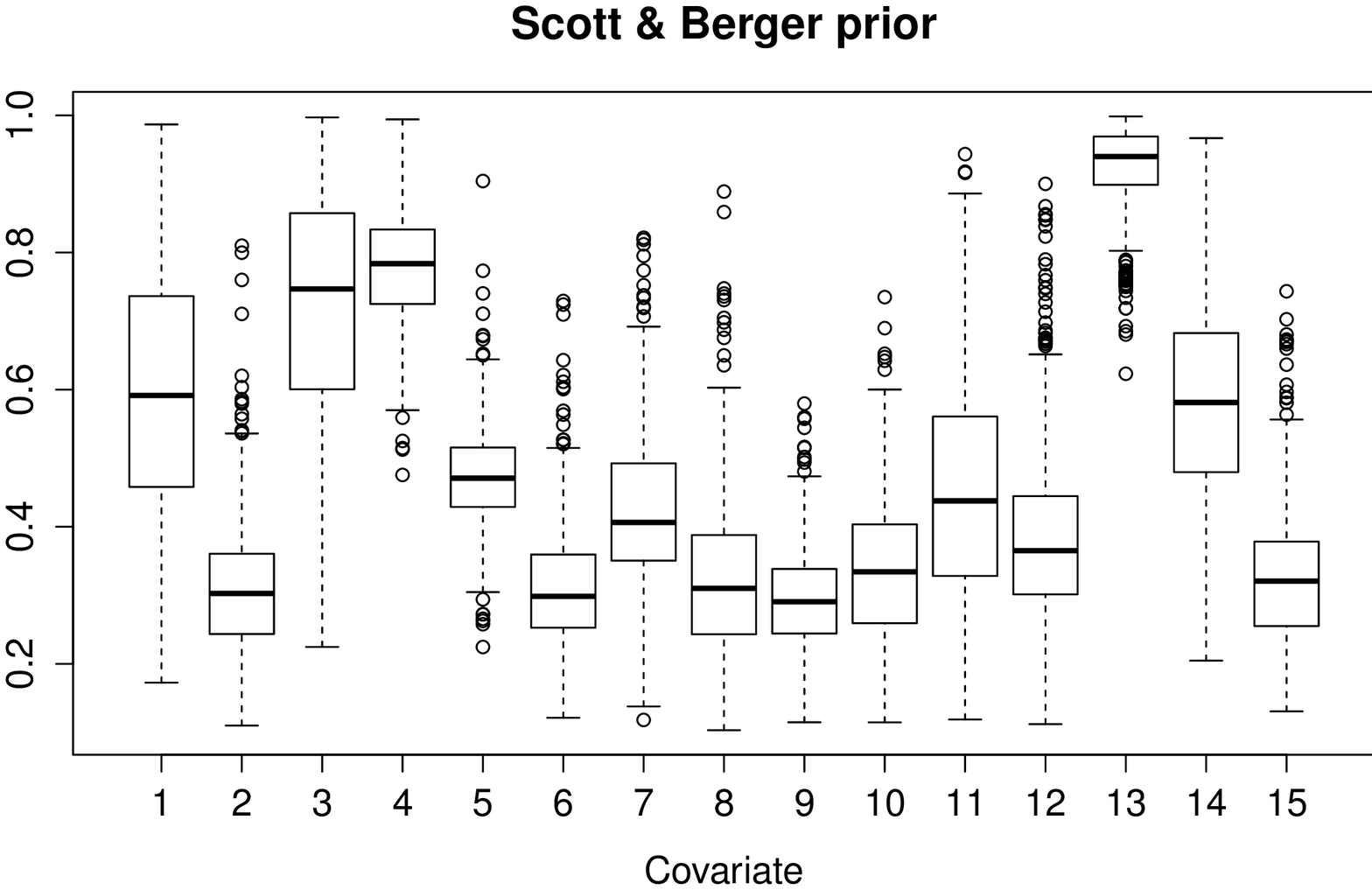} 
\label{US_RA_box:subfigur2}}
\subfigure[]{%
\includegraphics[scale=0.30]{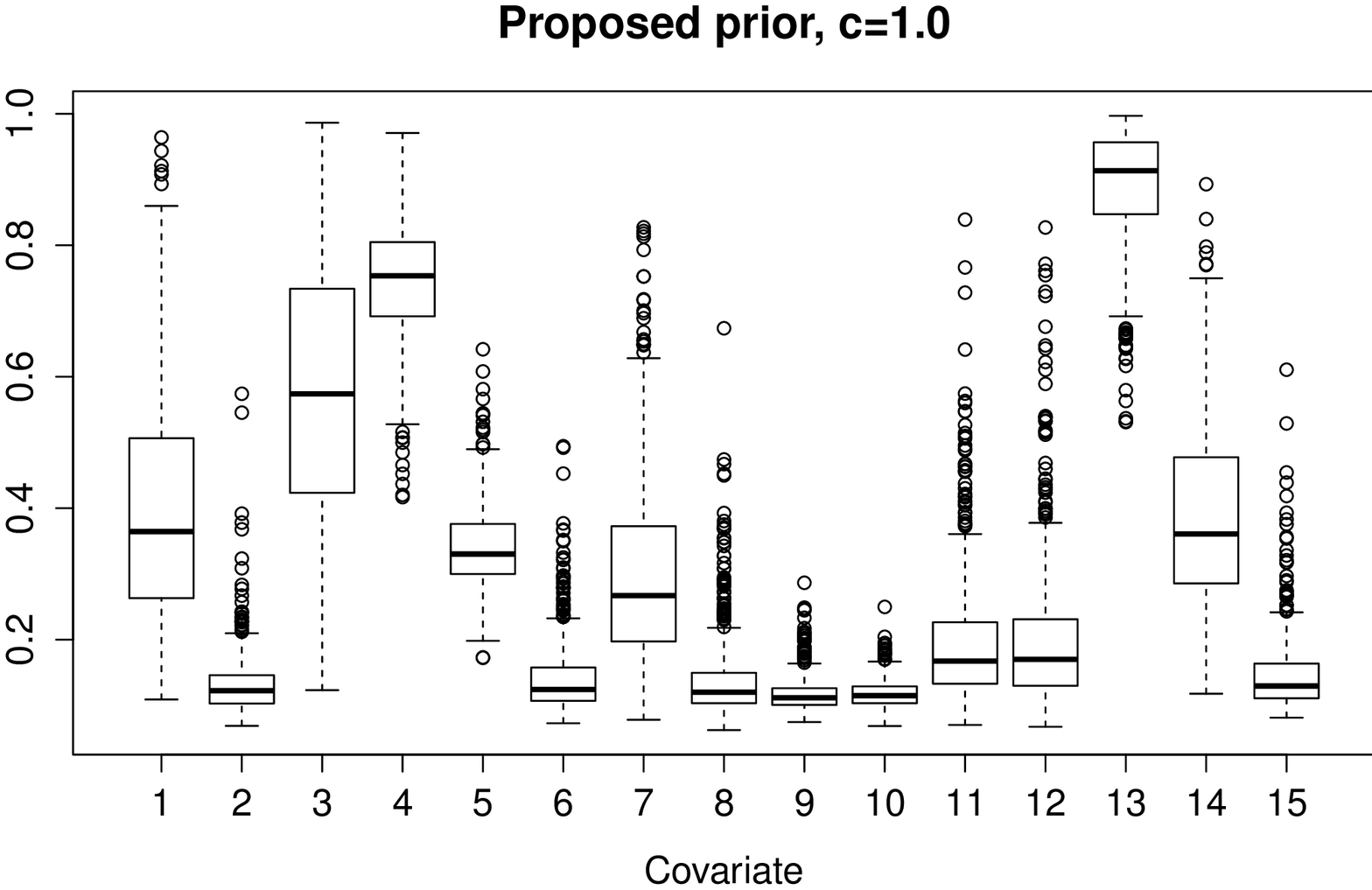}  
\label{US_RA_box:subfigur3}}
\quad
\subfigure[]{%
\includegraphics[scale=0.30]{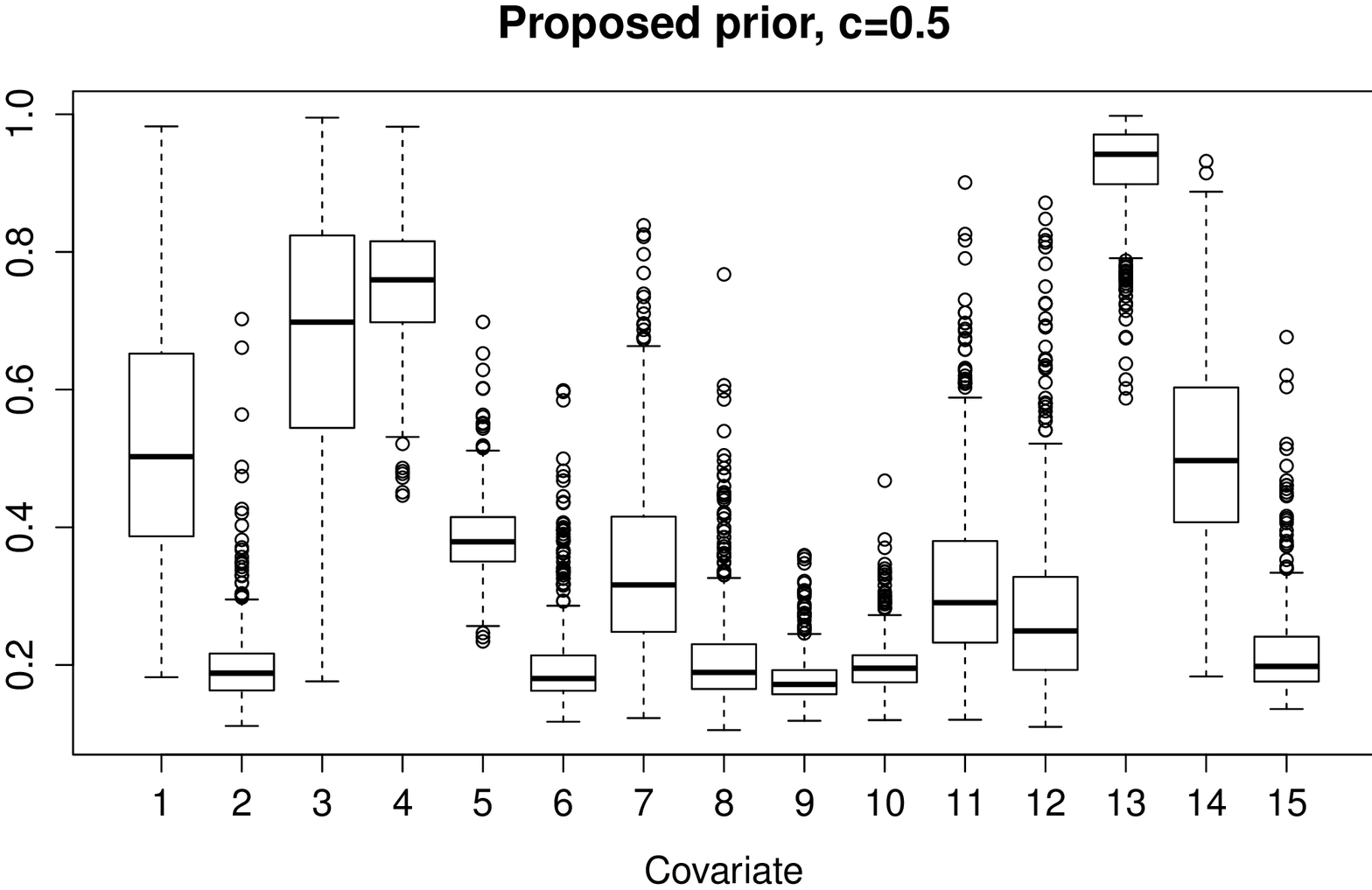} 
\label{US_RA_box:subfigur4}}
\subfigure[]{%
\includegraphics[scale=0.30]{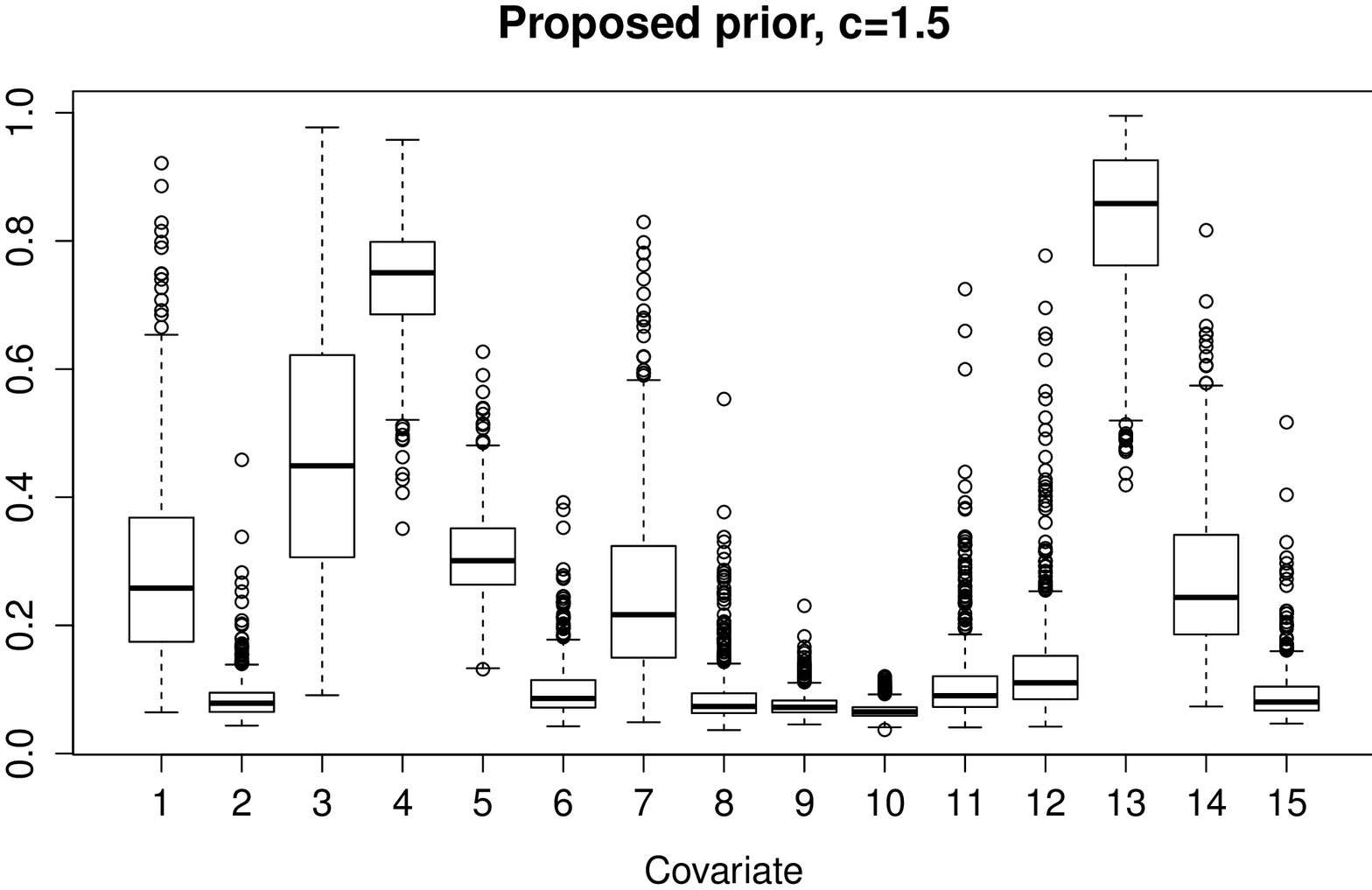}  
\label{US_RA_box:subfigur5}}
\quad
\subfigure[]{%
\includegraphics[scale=0.30]{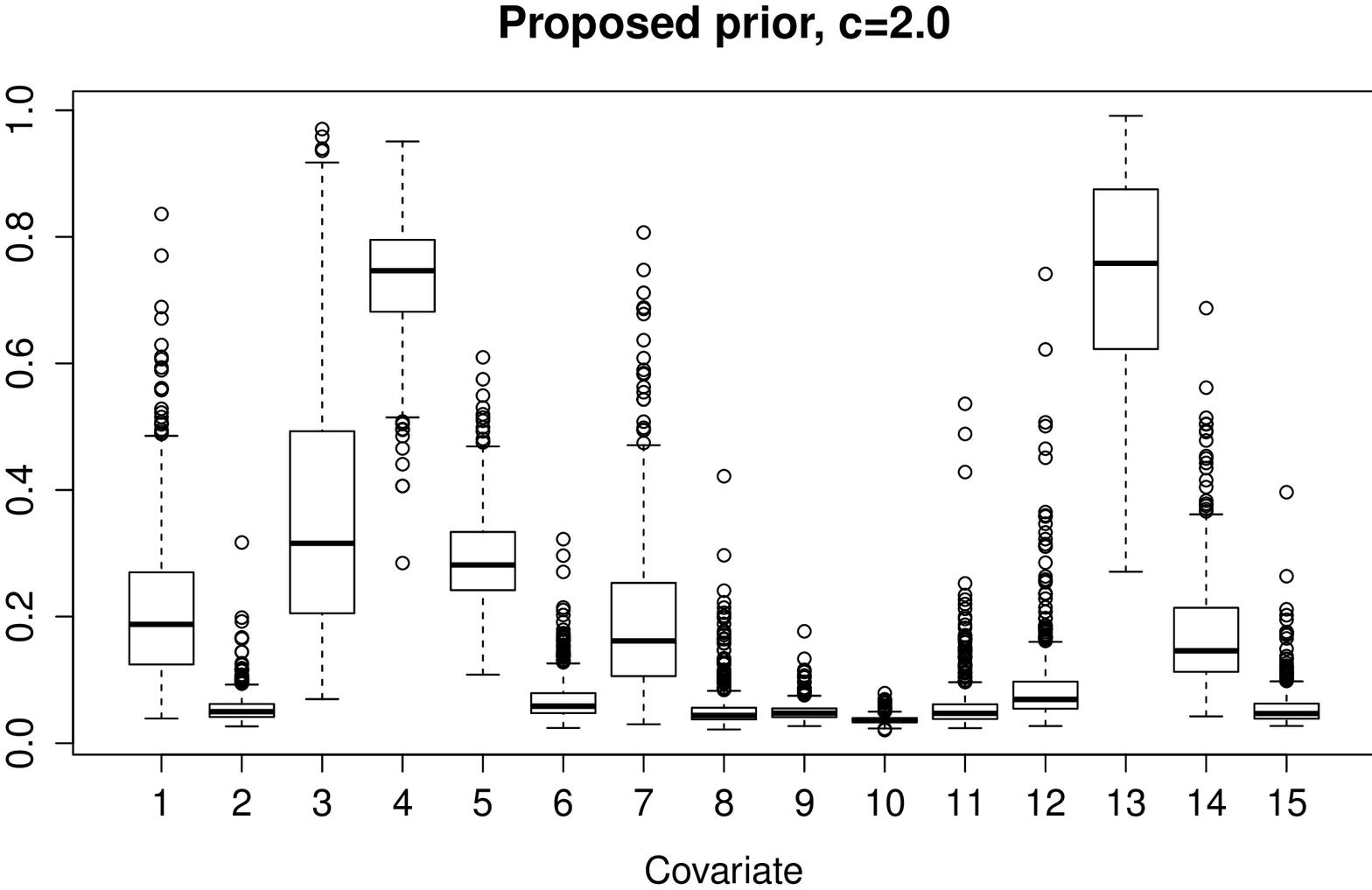}  
\label{US_RA_box:subfigur6}}
\caption{Box-plots of the posterior inclusion probabilities for the considered priors, where the proposed prior has been analysed for $c=1.0$, $c=0.5$, $c=1.5$ and $c=2.0$. Results are based on 500 replicates of sub-samples size of 40 from the US Crime data set.}
\label{fig:US_RA_box}
\end{figure}

For what it concerns the posterior inclusion probabilities, in Figure \ref{fig:US_RA_box}, we see a relatively small impact of $c$ on the variability of the probabilities, except when $c = 2.0$ when we see a larger variability for covariate number 13, for example. Effects of different values of the constant $c$ appear in the mean of the estimates. For example, covariates 1 and 3. When we compare our prior with $c=1$, to the uniform prior and, in particular, Scott \& Berger's, it is interesting to focus on the three covariates that are included in the MPM under these two last priors but not under our prior. In Figure \ref{fig:US_RA_box} these covariates are indicated by numbers 1, 11 and 14. We note that the size of the variability of their posterior inclusion probabilities under the proposed prior admits possible inclusion of the covariates in the MPM. This seems to suggest that, although our prior has undoubtedly a feature of estimating regression models with less covariates than the one defined by the other two priors, it possibly reflects a low level of certainty in the data on whether the covariates have to be considered or not. Considering the covariates included in the MPM, the covariates 3, 4 and 13 in Figure \ref{fig:US_RA_box}, we can make the following comments. For the covariates 4 and 13, there is little to argue as the distributions of the respective posterior inclusion probabilities show strong information in the data in favour of their inclusion in the MPM model. However, with similar arguments for the above non included covariates, the variability of the posterior inclusion probability of covariate 3 expresses a less strong support from the observations on whether the covariate has to be part of the regression model.

\begin{figure}[h!]
\centering
\subfigure[]{%
\includegraphics[scale=0.30]{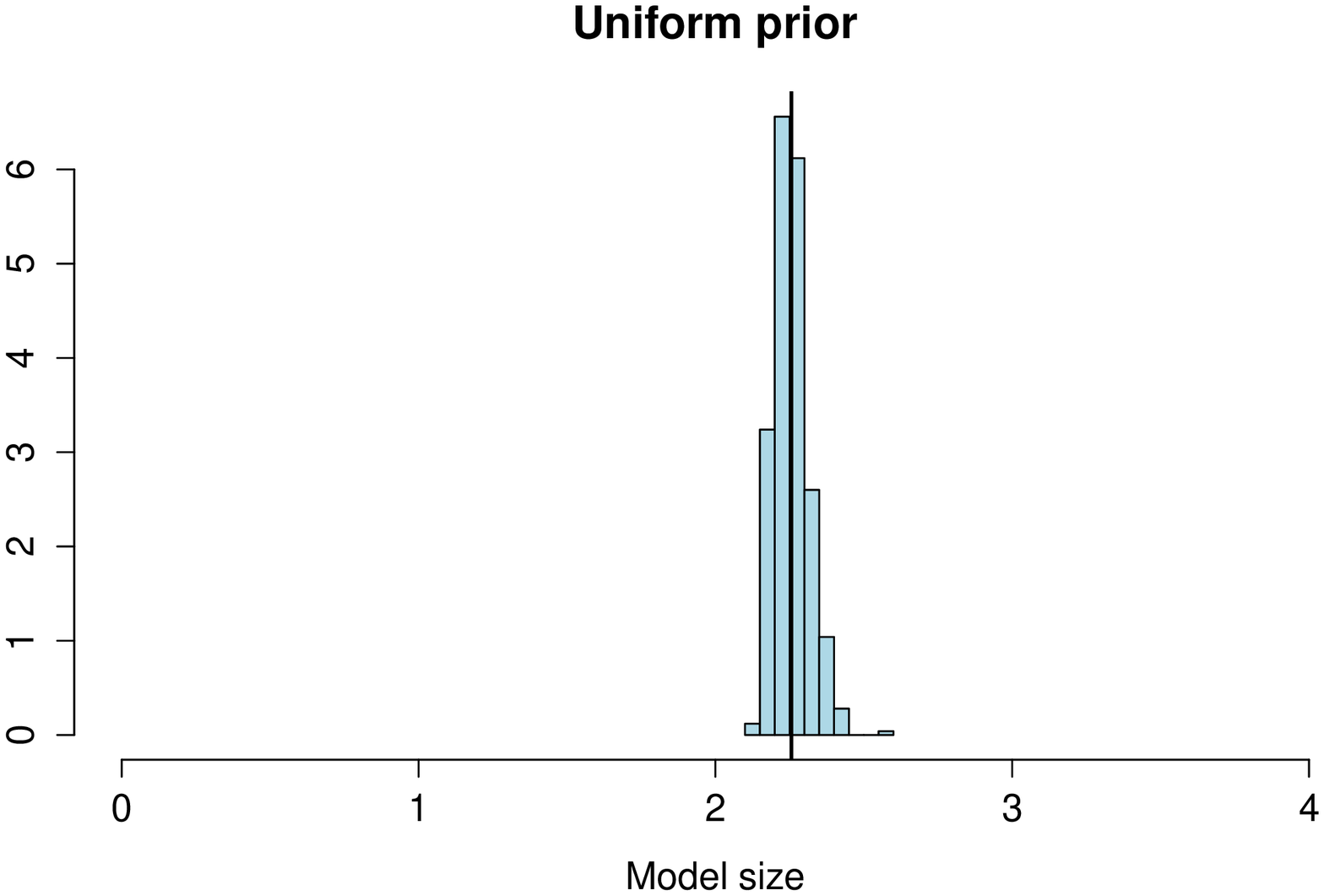}  
\label{Hald_RA_box:subfigur1}}
\quad
\subfigure[]{%
\includegraphics[scale=0.30]{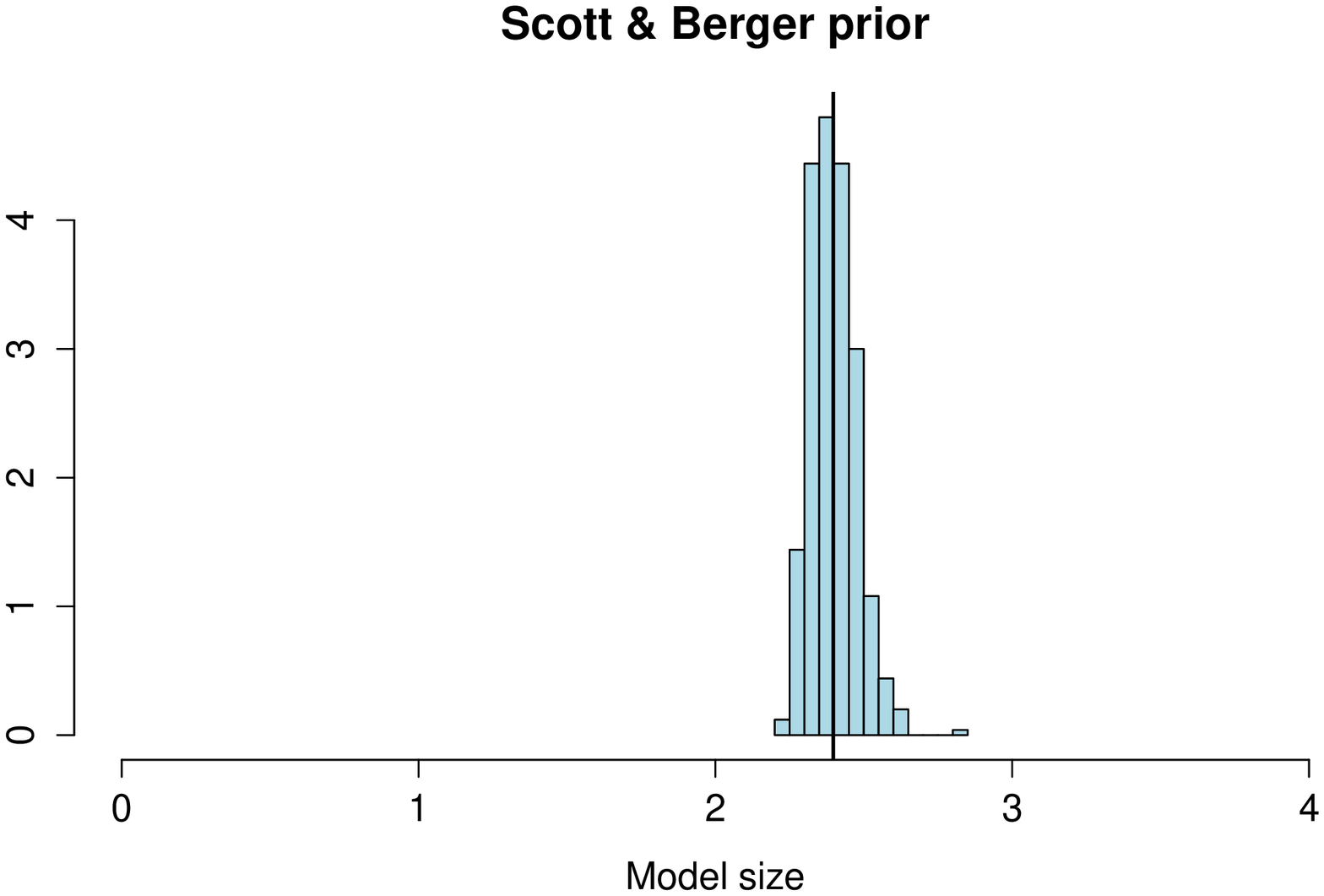} 
\label{Hald_RA_box:subfigur2}}
\subfigure[]{%
\includegraphics[scale=0.30]{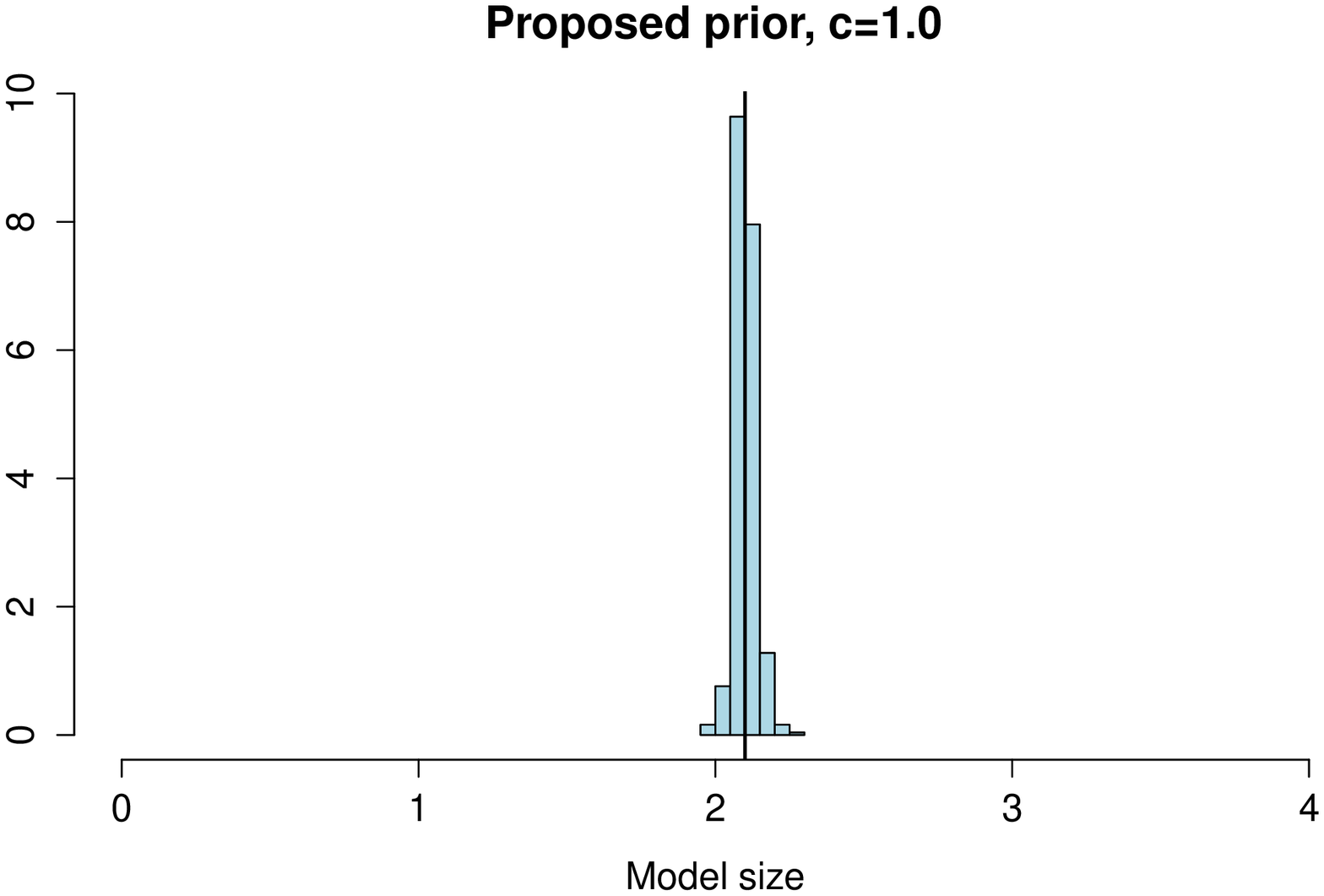}  
\label{Hald_RA_box:subfigur3}}
\quad
\subfigure[]{%
\includegraphics[scale=0.30]{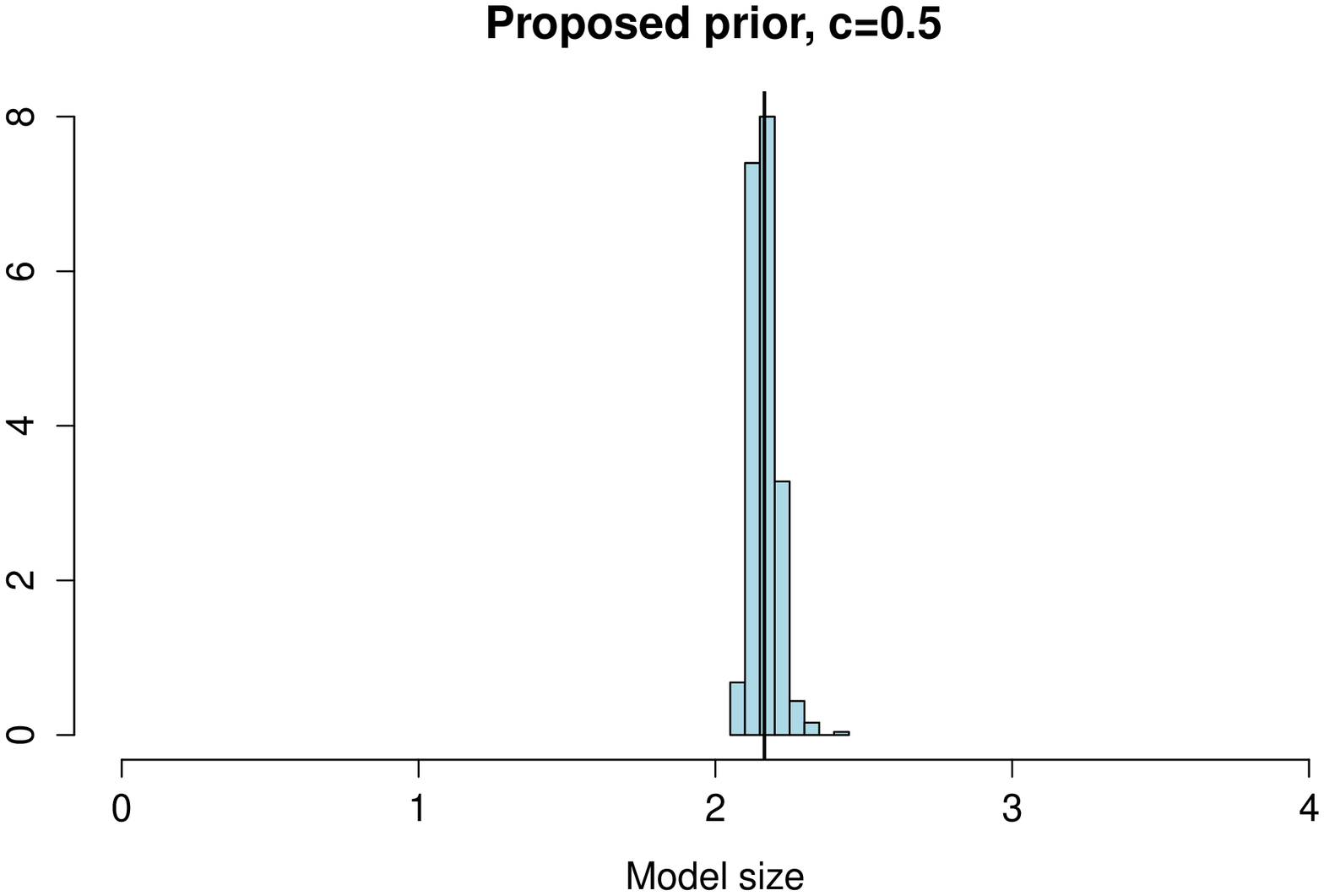} 
\label{Hald_RA_box:subfigur4}}
\subfigure[]{%
\includegraphics[scale=0.30]{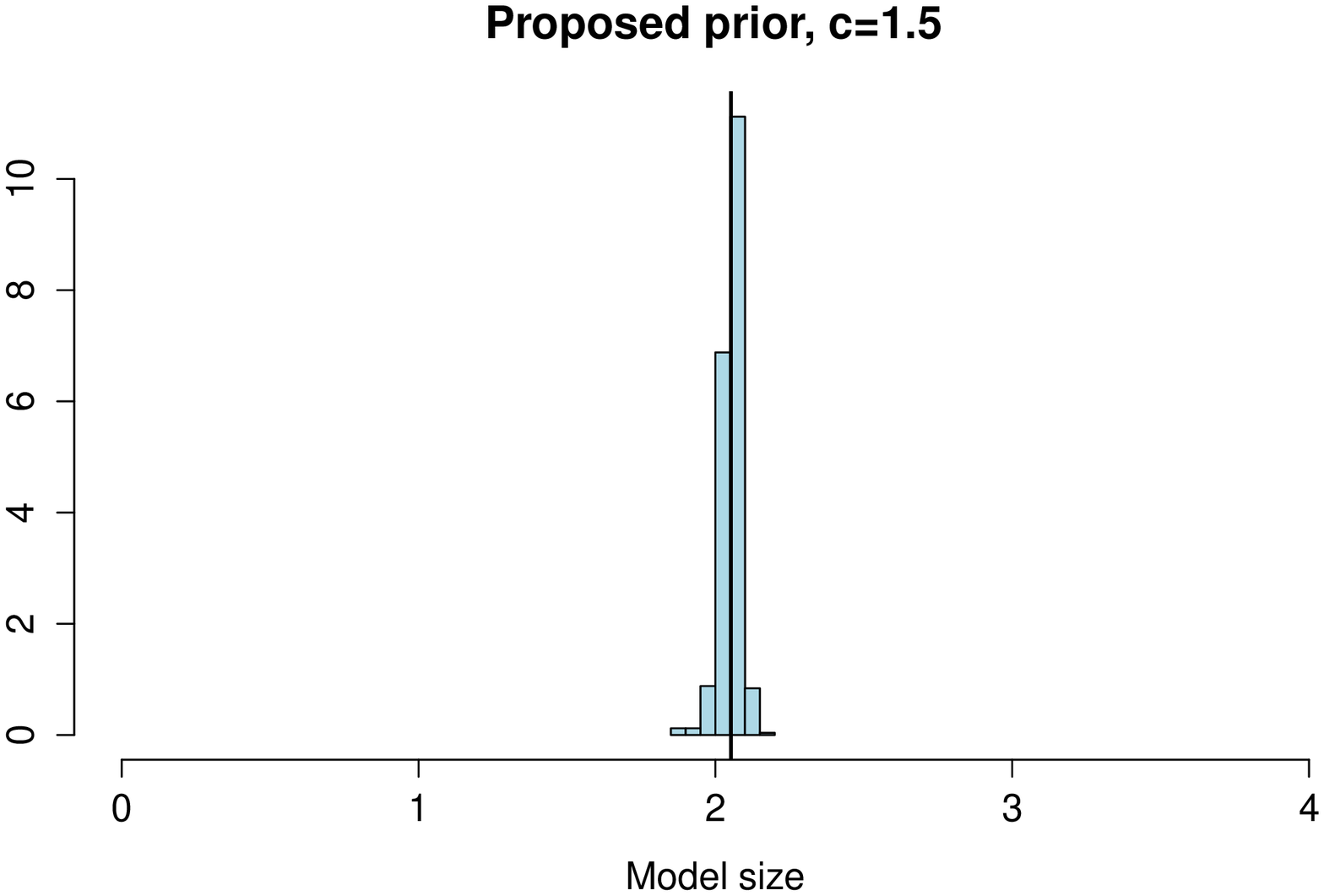}  
\label{Hald_RA_box:subfigur5}}
\quad
\subfigure[]{%
\includegraphics[scale=0.30]{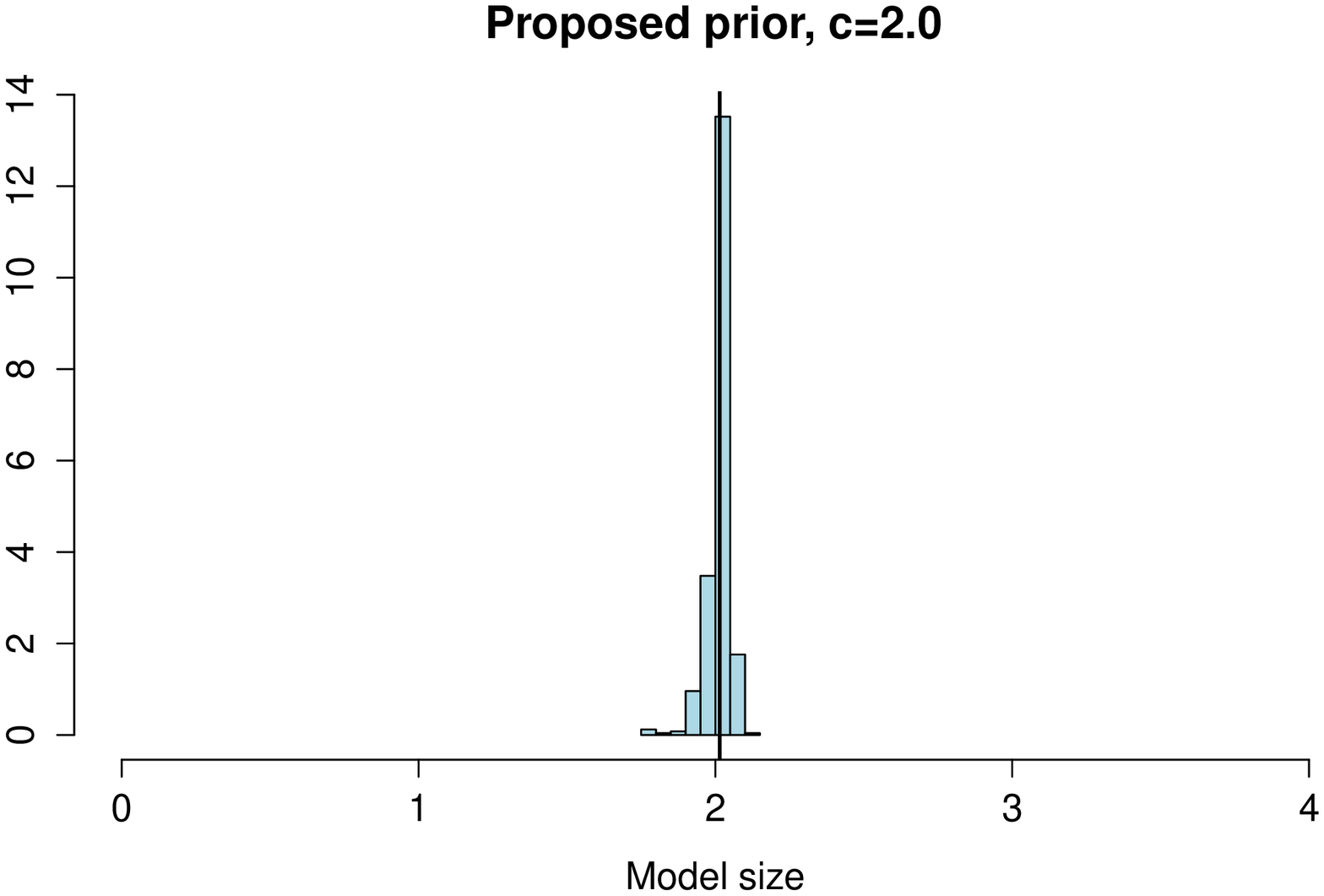}  
\label{Hald_RA_box:subfigur6}}
\caption{Histograms of the posterior mean model size for the considered priors, where the proposed prior has been analysed for $c = 1.0$, $c = 0.5$, $c = 1.5$ and $c = 2.0$ (500 sub-samples of size 10 from the Hald data set).}
\label{fig:Hald_RA_Hist}
\end{figure}

\begin{figure}[h!]
\centering
\subfigure[]{%
\includegraphics[scale=0.30]{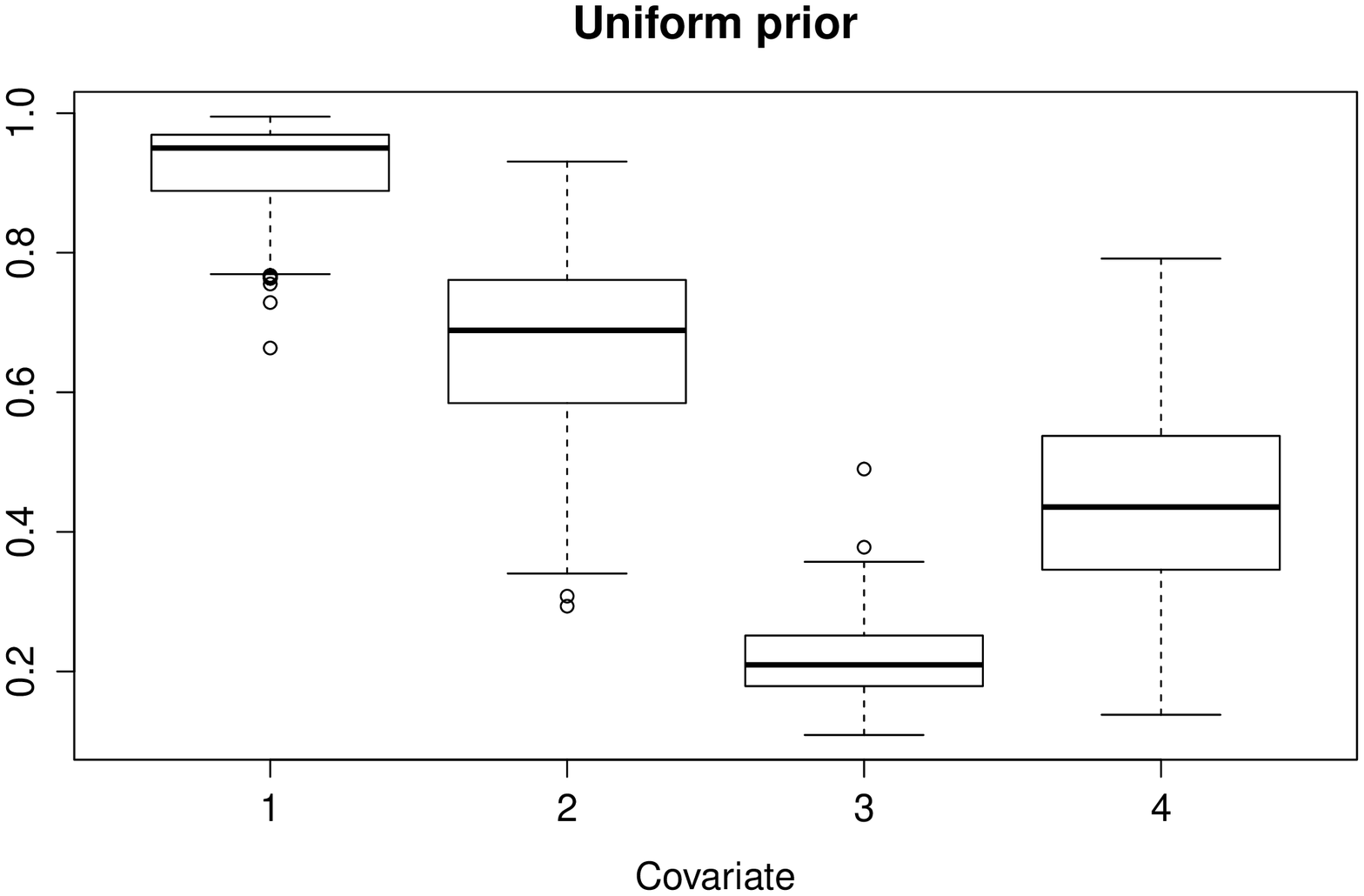}  
\label{Hald_RA_box:subfigur12}}
\quad
\subfigure[]{%
\includegraphics[scale=0.30]{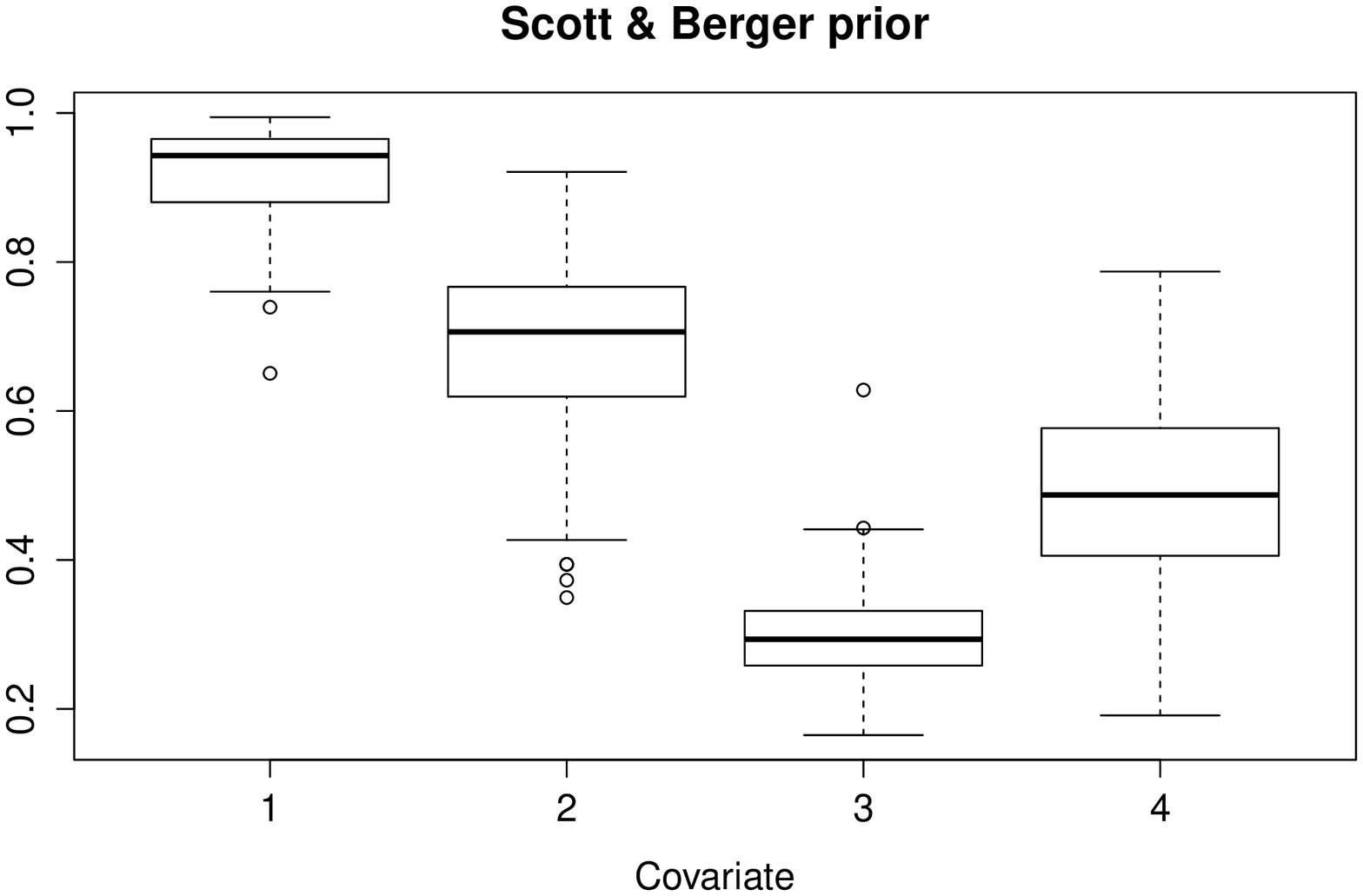} 
\label{Hald_RA_box:subfigur22}}
\subfigure[]{%
\includegraphics[scale=0.30]{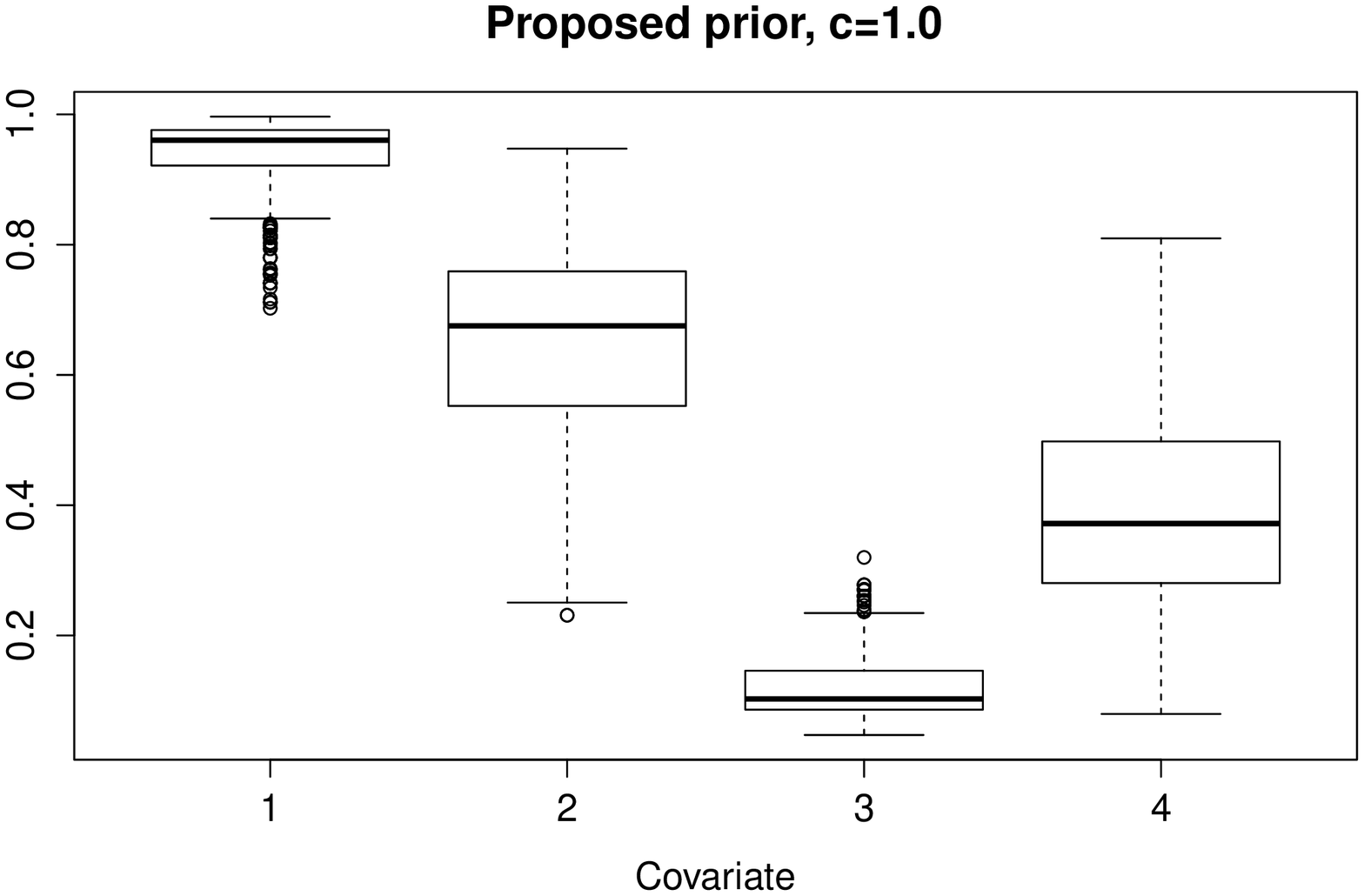}  
\label{Hald_RA_box:subfigur32}}
\quad
\subfigure[]{%
\includegraphics[scale=0.30]{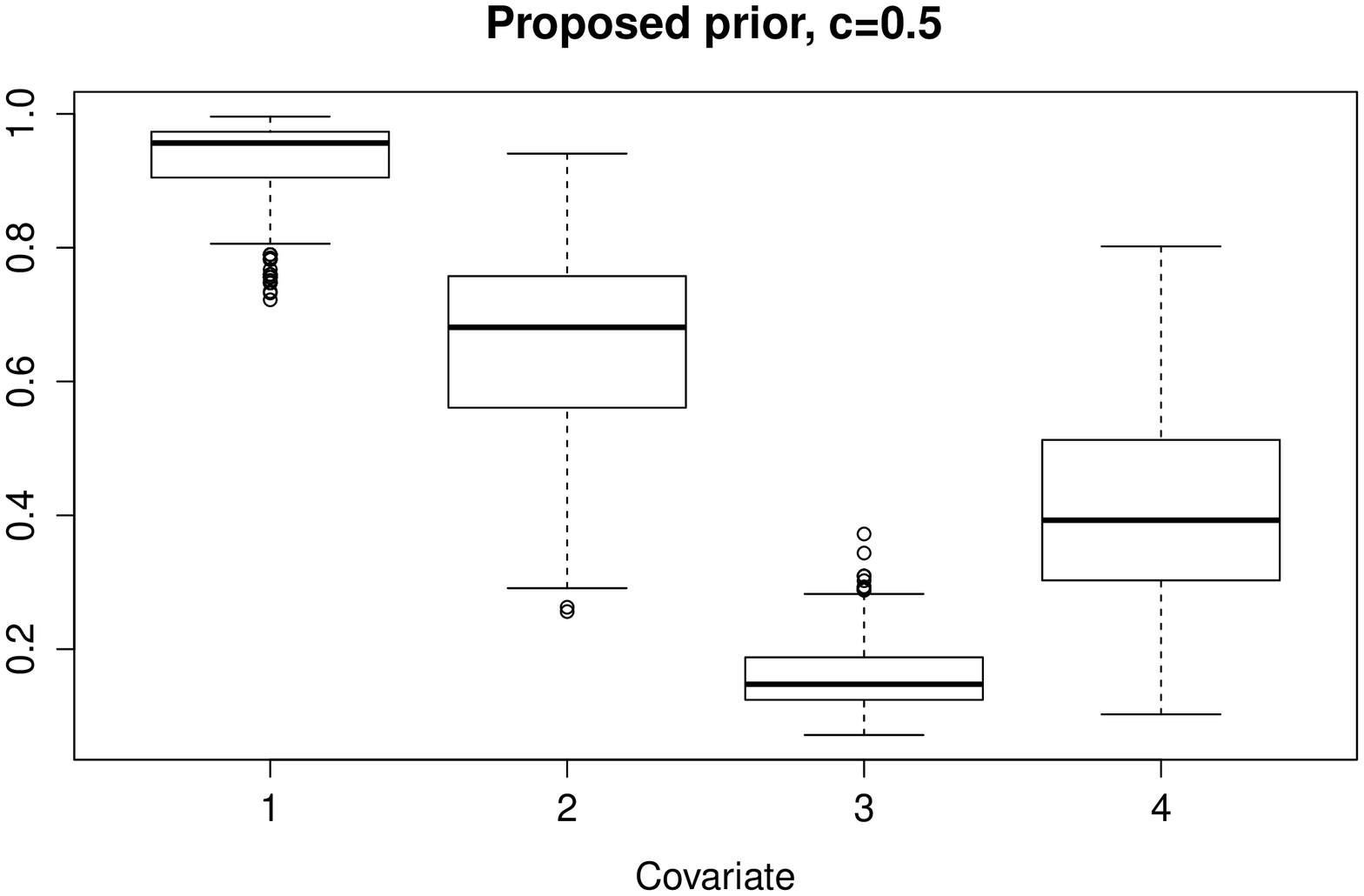} 
\label{Hald_RA_box:subfigur42}}
\subfigure[]{%
\includegraphics[scale=0.30]{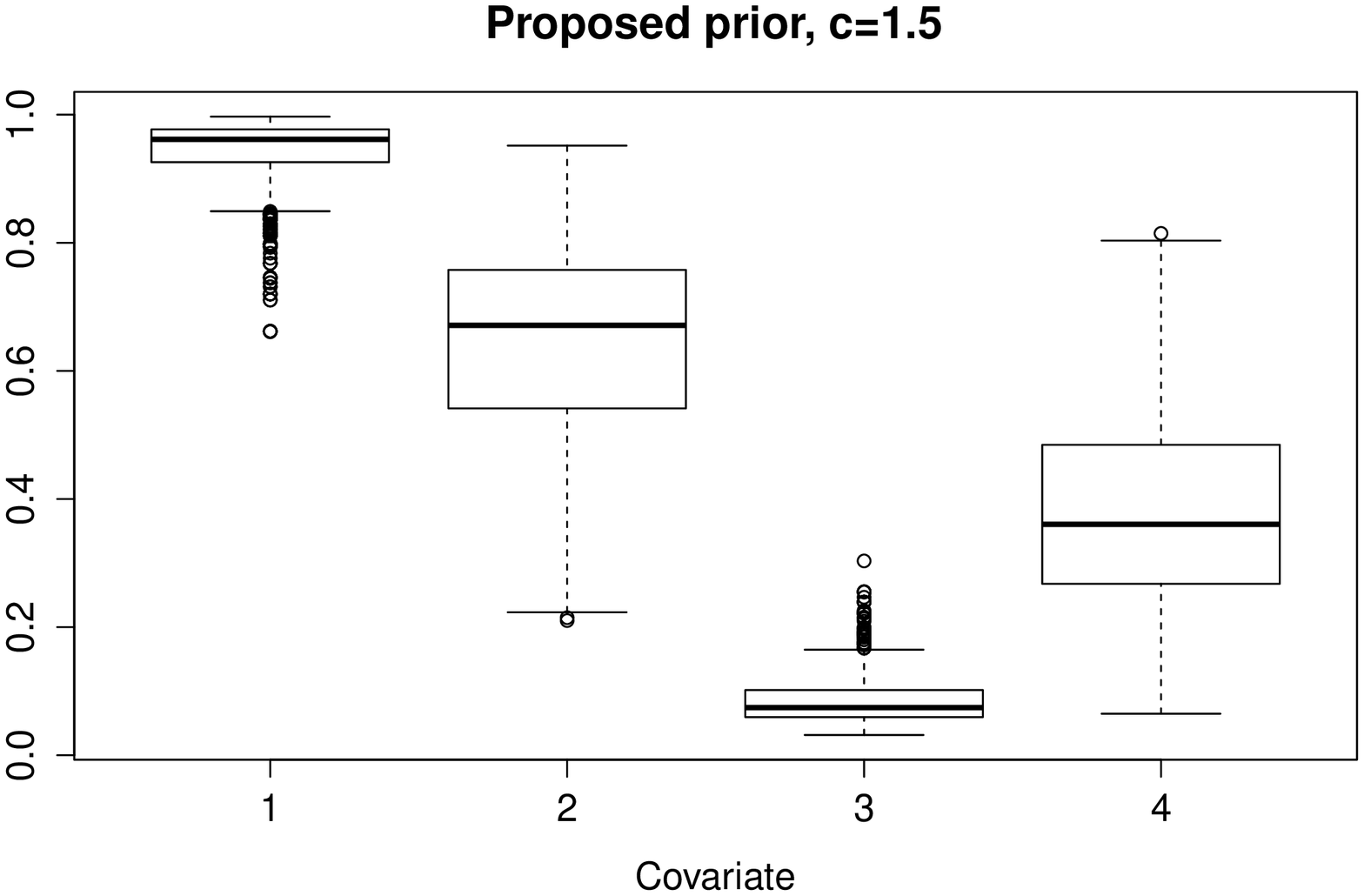}  
\label{Hald_RA_box:subfigur52}}
\quad
\subfigure[]{%
\includegraphics[scale=0.30]{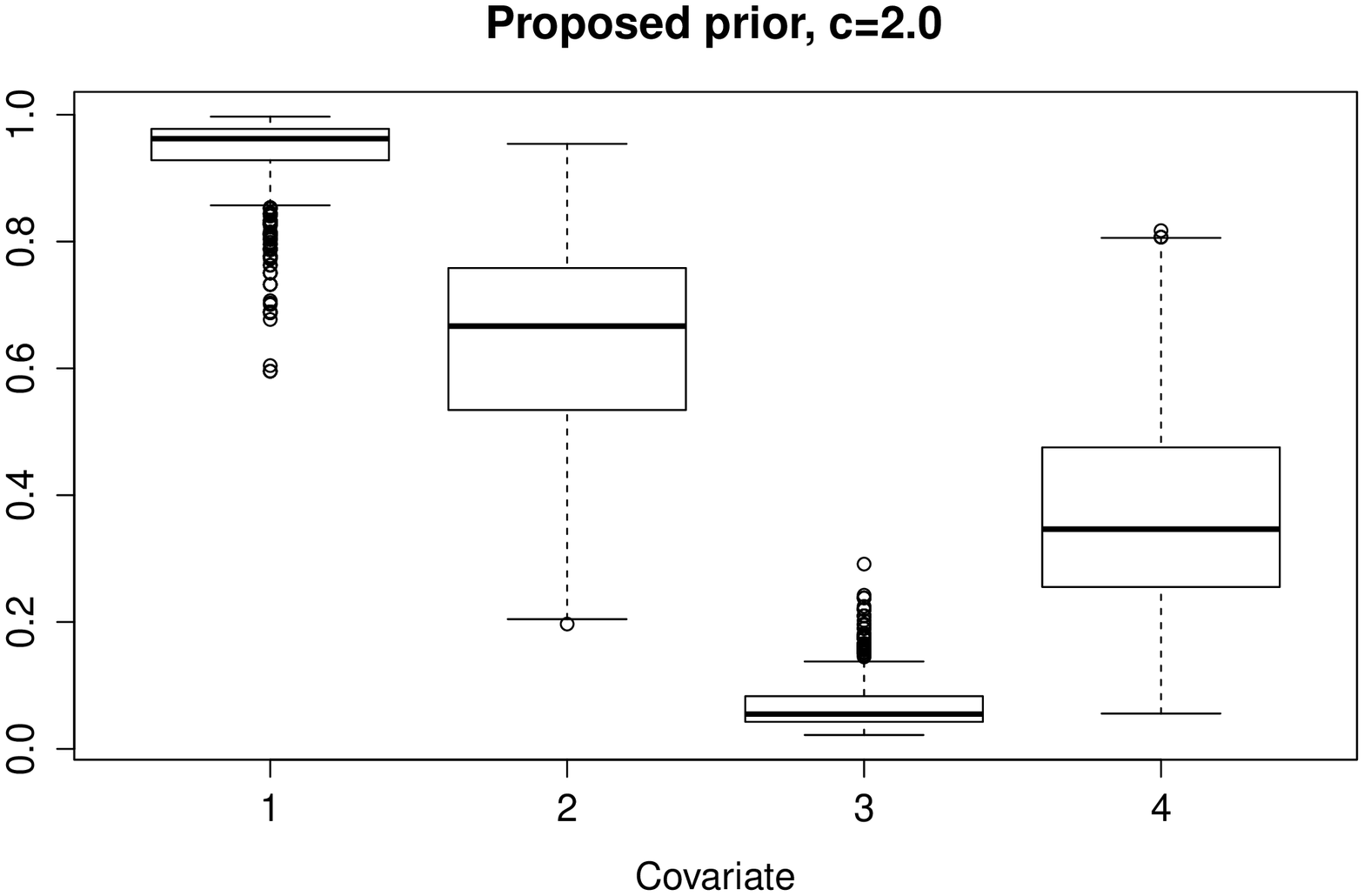} 
\label{Hald_RA_box:subfigur62}}
\caption{Box-plots of the posterior inclusion probabilities for the considered priors, where the proposed prior has been analysed for $c=1.0$, $c=0.5$, $c=1.5$ and $c=2.0$. Results are based on 500 replicates of sub-samples size of 10 from the Hald data set.}
\label{fig:Hald_RA_box}
\end{figure}

The robustness analysis of the Hald data set, with a similar procedure as above, consisted in 500 repeated samples of size 10. The histograms of the posterior mean model size are in Figure \ref{fig:Hald_RA_Hist}, while the box-plots of the posterior inclusion probabilities are in Figure \ref{fig:Hald_RA_box}. As already discussed in Section \ref{sc_hald}, the information about the covariates that should be part of the regression model contained in the data appears to be strong, and this is reflected in the robustness of the approach. In particular, the histograms of the posterior mean model size are concentrated around the mean obtained by considering the whole data set (vertical line).
The inspection of the box-plots of the posterior inclusion probabilities for this data set is straightforward. The impact of $c$ is more noticeable on the covariates with either a high posterior inclusion probability (i.e. number 1) or with a low posterior inclusion probability (i.e. number 3). In fact, for $c$ increasing, these probabilities become more prominent. In general, however, it doesn't seem that different priors lead to significantly different variability or shape of the distribution of the posterior inclusion probabilities.

%--- DISCUSSION ---------------------------------------------------------------------------------
\section{Discussion}\label{sc_conclusion}
This paper introduces a novel prior distribution for the model space in variable selection for linear regression. The prior is based on the idea that, if the ``wrong'' model is chosen, we incur in a loss that has two components: one represents the loss in information objectively measured by considering the \emph{worth} of the model, and one related to the complexity of the model expressed by its size. The obtained model has the following two appealing properties. First, it has a simple expression. Second, it has the flexibility of representing minimal prior information about the model as well as to allow the inclusion of appropriate prior information with regard to the model size.

The simulations studies carried out show good frequentist performance of the model prior when compared to other two existing minimally informative priors: the uniform prior and the Scott \& Berger prior \citep{SB10}. In particular, when we set the constant $c=1$, which is our recommendation for a prior carrying minimal prior information, the proposed prior outperforms the other two priors when the size of the true model is relatively small, with respect to the full model. In cases where the true model is close to the full model our prior is not performing as well as the other two priors. However, the difference in the performance is small and, if prior information is available, this lack of performance can be compensated by an appropriate choice of the constant $c$ reflecting the prior knowledge.

When it comes to real data analysis, we note a fair closeness of the results obtained by using the proposed prior with the ones of Scott \& Berger's. Both of them are more parsimonious than the uniform prior, unless the data itself is highly informative about which covariates should be included in the regression model. It is however important to highlight that the observed discrepancies between the proposed prior and the Scott \& Berger prior are limited to the median probability model. In fact, the estimated highest probability models are the same for both the analysed data sets, and it is noteworthy to mention that the posterior probability associated to the highest probability model under our prior is the largest.

%--- APPENDIX -----------------------------------------------------------------------------------
%\begin{appendices}
\section*{Appendix - Proof of Theorem \ref{teo_KLreg1}}

Let $\pmb\gamma$ and $\pmb\gamma^\prime$ be non-identical binary vectors. With the corresponding design matrices $\mathbf{X}_{\pmb\gamma}$ and  $\mathbf{X}_{\pmb\gamma^\prime}$ and the vectors of coefficients $\pmb\beta_{\pmb\gamma}$ and $\pmb\beta_{\pmb\gamma^\prime}$, the models are defined as 

\[ M_{\pmb\gamma} : \mathbf{y} \sim N_n\left( \mathbf{1}_n\alpha+ \mathbf{X}_{\pmb\gamma} \pmb\beta_{\pmb\gamma}, \mbox{I}/\phi\right) = N_n\left( \tilde{\mathbf{X}}_i \tilde{\pmb\beta}_{\pmb\gamma} , \mbox{I}/\phi\right) \]
\[ M_{\pmb\gamma^\prime} : \mathbf{y} \sim N_n \left( \mathbf{1}_n\alpha+ \mathbf{X}_{\pmb\gamma^\prime} \pmb\beta_{\pmb\gamma^\prime}, \mbox{I}/\phi\right) = N_n\left( \tilde{\mathbf{X}}_{\pmb\gamma^\prime} \tilde{\pmb\beta}_{\pmb\gamma^\prime}, \mbox{I}/\phi\right)  \]
where $\tilde{\mathbf{X}}_{\pmb\gamma}=[\mathbf{1},\mathbf{X}_{\pmb\gamma}]$, $\tilde{\mathbf{X}}_{\pmb\gamma^\prime}=[\mathbf{1},\mathbf{X}_{\pmb\gamma^\prime}]$, $\tilde{\pmb\beta}_{\pmb\gamma}=(\alpha,\pmb\beta_{\pmb\gamma})^T$, and $\tilde{\pmb\beta}_{\pmb\gamma^\prime}=(\alpha,\pmb\beta_{\pmb\gamma^\prime})^T$. 
The Kullback--Leibler divergence of $M_{\pmb\gamma}$ from $M_{\pmb\gamma^\prime}$ is
\[\begin{array}{c l}
&D_{KL}(M_{\pmb\gamma}\|M_{\pmb\gamma^\prime}) \\ =& \displaystyle\int m(\mathbf{y}|M_{\pmb\gamma}) \Bigg \{ -\dfrac{\log(1/\phi)}{2}-\dfrac{\phi}{2}(\mathbf{y}-\tilde{\mathbf{X}}_{\pmb\gamma} \tilde{\pmb\beta}_{\pmb\gamma})^T(\mathbf{y}-\tilde{\mathbf{X}}_{\pmb\gamma} \tilde{\pmb\beta}_{\pmb\gamma})  + \dfrac{\log(1/\phi)}{2} \\ & +\dfrac{\phi}{2}(\mathbf{y}-\tilde{\mathbf{X}}_{\pmb\gamma^\prime} \tilde{\pmb\beta}_{\pmb\gamma^\prime})^T(\mathbf{y}-\tilde{\mathbf{X}}_{\pmb\gamma^\prime} \tilde{\pmb\beta}_{\pmb\gamma^\prime}) \Bigg \}\, d\mathbf{y} \\
=&\displaystyle\int m(\mathbf{y}|M_{\pmb\gamma}) \phi \Bigg \{ -\dfrac{1}{2}\mathbf{y}^T\mathbf{y}+(\tilde{\mathbf{X}}_{\pmb\gamma}\tilde{\pmb\beta}_{\pmb\gamma})^T\mathbf{y}-\dfrac{1}{2}(\tilde{\mathbf{X}}_{\pmb\gamma}\tilde{\pmb\beta}_{\pmb\gamma})^T(\tilde{\mathbf{X}}_{\pmb\gamma}\tilde{\pmb\beta}_{\pmb\gamma}) + \dfrac{1}{2}\mathbf{y}^T\mathbf{y}-(\tilde{\mathbf{X}}_{\pmb\gamma^\prime}\tilde{\pmb\beta}_{\pmb\gamma^\prime})^T\mathbf{y} \\ & +\dfrac{1}{2}(\tilde{\mathbf{X}}_{\pmb\gamma^\prime}\tilde{\pmb\beta}_{\pmb\gamma^\prime})^T(\tilde{\mathbf{X}}_{\pmb\gamma^\prime}\tilde{\pmb\beta}_{\pmb\gamma^\prime}) \Bigg\} \, d\mathbf{y} \\
=& \dfrac{\phi}{2}\Big ( (\tilde{\mathbf{X}}_{\pmb\gamma}\tilde{\pmb\beta}_{\pmb\gamma})^T(\tilde{\mathbf{X}}_{\pmb\gamma}\tilde{\pmb\beta}_{\pmb\gamma})
-2(\tilde{\mathbf{X}}_{\pmb\gamma^\prime}\tilde{\pmb\beta}_{\pmb\gamma^\prime})^T(\tilde{\mathbf{X}}_{\pmb\gamma}\tilde{\pmb\beta}_{\pmb\gamma}) + (\tilde{\mathbf{X}}_{\pmb\gamma^\prime}\tilde{\pmb\beta}_{\pmb\gamma^\prime})^T(\tilde{\mathbf{X}}_{\pmb\gamma^\prime}\tilde{\pmb\beta}_{\pmb\gamma^\prime}) 
\Big) \end{array}\]

The first derivative of $D_{KL}(M_{\pmb\gamma}\|M_{\pmb\gamma^\prime}) $ with respect to $\tilde{\pmb\beta}_{\pmb\gamma^\prime}$ is 
\[ \dfrac{\partial D_{KL}(M_{\pmb\gamma}\|M_{\pmb\gamma^\prime}) }{\partial\tilde{\pmb\beta}_{\pmb\gamma^\prime}} = \phi\Big(-(\tilde{\mathbf{X}}_{\pmb\gamma}\tilde{\pmb\beta}_{\pmb\gamma})^T +(\tilde{\mathbf{X}}_{\pmb\gamma^\prime}\tilde{\pmb\beta}_{\pmb\gamma^\prime})^T \Big) \tilde{\mathbf{X}}_{\pmb\gamma^\prime}.\]
If $\tilde{\mathbf{X}}_{\pmb\gamma^\prime}\tilde{\mathbf{X}}_{\pmb\gamma^\prime}^T$ is invertible, $D_{KL}(M_{\pmb\gamma}\|M_{\pmb\gamma^\prime}) $ is minimized when $\tilde{\pmb\beta}_{\pmb\gamma^\prime}=(\tilde{\mathbf{X}}^T_{\pmb\gamma^\prime}\tilde{\mathbf{X}}_{\pmb\gamma^\prime})^{-1}\tilde{\mathbf{X}}_{\pmb\gamma^\prime}^T\tilde{\mathbf{X}}_{\pmb\gamma}\tilde{\pmb\beta}_{\pmb\gamma} $ and the minimum is given by 
\[ \min_{\tilde{\pmb\beta}_{\pmb\gamma^\prime}} D_{KL}(M_{\pmb\gamma}\|M_{\pmb\gamma^\prime})=0. \]

%------------------------------------------------------------------------------------------------

\end{document}